\documentclass[acmsmall,authorversion,nonacm]{acmart}

\usepackage[linesnumbered,ruled,vlined]{algorithm2e}
\usepackage[export]{adjustbox}
\usepackage{amsmath}
\usepackage{array}
\usepackage{arydshln}
\usepackage{bbm}
\usepackage{bm}
\usepackage{caption}
\usepackage{float}
\usepackage{graphicx}
\usepackage{listings}
\usepackage{longtable}
\usepackage{mathpartir}
\usepackage{multirow}
\usepackage{enumitem}
\usepackage{pifont}
\usepackage{pgfplots}
\usepackage{relsize}
\usepackage{subcaption}
\usepackage{stackengine}
\usepackage{stmaryrd}
\usepackage{tablefootnote}
\usepackage{tabularray}
\usepackage{tcolorbox}
\usepackage{tikz, pgfplots}
\usepackage{wrapfig}
\usepackage{verbatimbox}
\usepackage{pgfplotstable}
\usetikzlibrary{patterns}
\usetikzlibrary{positioning}

\newcommand{\ours}{Scallop}
\newcommand{\ourpybind}{\texttt{scallopy}}
\newcommand{\ourwasmbind}{\texttt{scallop-wasm}}

\newcommand{\sep}[0]{~|~}
\newcommand{\sem}[1]{\ensuremath{\llbracket #1 \rrbracket}}
\newcommand{\ram}[0]{{\sc SclRam}}

\newcommand{\code}[1]{\ensuremath{\texttt{#1}}}
\newcommand{\inlinecodesmall}[1]{{\texttt{\small #1}}}
\newcommand{\inlinecode}[1]{{\texttt{\footnotesize #1}}}
\newcommand{\inlinescl}[1]{{\footnotesize \lstinline[language=scallop]!#1!}}

\newcommand{\contains}[0]{~\scalebox{1.2}{\ensuremath{\vDash}}~}

\newcommand{\anno}[1]{{\footnotesize (#1)}}
\newcommand{\old}{{\texttt{old}}}
\newcommand{\new}{{\texttt{new}}}

\newcommand{\tick}{\ding{52}}

\makeatletter
\newcommand{\oeq}{\mathbin{\mathpalette\make@circled{=}}}
\newcommand{\oneq}{\mathbin{\mathpalette\make@circled{\neq}}}
\newcommand{\make@circled}[2]{%
  \ooalign{$\m@th#1\smallbigcirc{#1}$\cr\hidewidth$\m@th#1#2$\hidewidth\cr}%
}
\newcommand{\smallbigcirc}[1]{%
  \vcenter{\hbox{\scalebox{0.73}{$\m@th#1\bigcirc$}}}%
}
\makeatother

\definecolor{sclgreen}{rgb}{0,0.46,0}
\definecolor{sclblue}{rgb}{0.02,0.42,0.74}
\definecolor{scllightgrey}{rgb}{0.94,0.94,0.94}%
\definecolor{sclgreyblue}{rgb}{0.3,0.4,0.6}%
\definecolor{sclcyan}{rgb}{0.1,0.4,0.6}%
\definecolor{sclpurple}{rgb}{0.71,0,0.85}
\definecolor{sclyellow}{rgb}{0.9,0.6,0.05}
\definecolor{sclorange}{rgb}{1,0.36,0.03}
\definecolor{sclred}{rgb}{0.6,0.2,0.0}%

\SetCommentSty{mycommfont}

\lstdefinelanguage{scallop}{
    keywords={import,type,const,rel,query,usize,where,as,String,i8,i32,i64,usize,u8,u16,u32,u64},keywordstyle=\color{blue},%
    morekeywords=[2]{and,or,not,implies,==,+,-,*,/},keywordstyle=[2]\color{sclpurple},%
    morekeywords=[3]{count,sum,prod,min,max,exists,forall,unique,top,categorical,uniform},keywordstyle=[3]\color{sclorange},%
    morecomment=[s]{/*}{*/},%
    commentstyle=\color{sclgreen},%
    morecomment=[l]{//},%
    morestring=[b]",stringstyle=\color{sclyellow}
}

\lstdefinelanguage{mypython}{
    keywords={class,def,str,return,if,elif,else,for,in,while,int,List,Tuple},keywordstyle=\color{blue},%
    morekeywords=[2]{self},
    keywordstyle=[2]\color{sclred},
    morekeywords=[3]{__init__},
    keywordstyle=[3]\color{sclcyan},
    morecomment=[s]{"""}{"""},commentstyle=\color{sclgreen},%
    morecomment=[l]{\#},%
    morestring=[b]",stringstyle=\color{sclorange}
}

\SetKwInput{KwInput}{Input}
\SetKwInput{KwOutput}{Output}

\newcommand{\secref}[1]{Section~\ref{#1}}

\newcommand{\figref}[1]{Fig.~\ref{#1}}
\newcommand{\tabref}[1]{Table~\ref{#1}}

\definecolor{mygreen}{HTML}{D5E8D4}
\definecolor{myred}{HTML}{F8CECC}
\definecolor{myorange}{HTML}{ffcf99}
\definecolor{myblue}{HTML}{99c8f2}
\definecolor{mypurple}{HTML}{E1D5E7}
\definecolor{myyellow}{HTML}{FFF2CC}

\definecolor{mydeepgreen}{HTML}{82B366}
\definecolor{mydeepred}{HTML}{B85450}
\definecolor{mydeeporange}{HTML}{e38820}
\definecolor{mydeepblue}{HTML}{408bcf}
\definecolor{mydeeppurple}{HTML}{9673A6}
\definecolor{mydeepyellow}{HTML}{D6B656}

\definecolor{mylightblue}{HTML}{cfe2fa}
\definecolor{mylightorange}{HTML}{faeccf}

\newcommand*{\hwfsample}[1]{%
    \raisebox{-.1\baselineskip}{%
        \includegraphics[
        height=0.7\baselineskip,
        keepaspectratio,
        ]{#1}%
    }%
}

\startPage{1}

\setcopyright{none}
\acmPrice{}
\acmDOI{10.1145/3591280}
\acmYear{2023}
\copyrightyear{2023}
\acmSubmissionID{pldi23main-p360-p}
\acmJournal{PACMPL}
\acmVolume{7}
\acmNumber{PLDI}
\acmMonth{6}

\usepackage{booktabs}   
\usepackage{subcaption} 

\begin{document}

\title{\ours: A Language for Neurosymbolic Programming}


\author{Ziyang Li}
\authornote{The two authors contributed equally on the paper}
\affiliation{
  \institution{University of Pennsylvania}
}
\email{liby99@seas.upenn.edu}

\author{Jiani Huang}
\authornotemark[1]
\affiliation{
  \institution{University of Pennsylvania}
}
\email{jianih@seas.upenn.edu}

\author{Mayur Naik}
\affiliation{
  \institution{University of Pennsylvania}
}
\email{mnaik@seas.upenn.edu}

\begin{abstract}
We present \ours, a language which combines the benefits of deep learning and logical reasoning.
\ours~ enables users to write a wide range of neurosymbolic applications and train them in a data- and compute-efficient manner.
It achieves these goals through three key features:
1) a flexible symbolic representation that is based on the relational data model;
2) a declarative logic programming language that is based on Datalog and supports recursion, aggregation, and negation; and
3) a framework for automatic and efficient differentiable reasoning that is based on the theory of provenance semirings.
We evaluate \ours~ on a suite of eight neurosymbolic applications from the literature.
Our evaluation demonstrates that \ours~ is capable of expressing algorithmic reasoning in diverse and challenging AI tasks, provides a succinct interface for machine learning programmers to integrate logical domain knowledge, and yields solutions that are comparable or superior to state-of-the-art models in terms of accuracy.
Furthermore, \ours's solutions outperform these models in aspects such as runtime and data efficiency, interpretability, and generalizability.

\end{abstract}



\maketitle

\section{Introduction}
\label{sec:introduction}

Classical algorithms and deep learning embody two prevalent paradigms of modern programming.
Classical algorithms are well suited for exactly-defined tasks, such as sorting a list of numbers or finding a shortest path in a graph.
Deep learning, on the other hand, is well suited for tasks that are not tractable or feasible to perform using classical algorithms, such as detecting objects in an image or parsing natural language text.
These tasks are typically specified using a set of input-output training data, and solving them involves learning the parameters of a deep neural network to fit the data using gradient-based methods.

\begin{figure}
  \footnotesize
  \begin{subfigure}[b]{0.25\textwidth}
    \centering
    \begin{tikzpicture}
      \node[draw,double,minimum height=0.45cm] (r) at (0,0) {$r$};
      \node[draw,minimum height=0.45cm] (p) [right=0.5cm of r] {$P$};
      \node[draw,double,minimum height=0.45cm] (y) [right=0.5cm of p] {$y$};
      \draw[->] (r.east) -- (p.west);
      \draw[->] (p.east) -- (y.west);
      \draw[->, white] ([yshift=-0.05cm]y.south) -- ([yshift=-0.4cm]y.south) -- ([yshift=-0.4cm]p.south) node [midway] {$\phantom{\frac{\partial}{\partial}}$} -- (p.south);
    \end{tikzpicture}
    \vspace{-0.07in}
    \caption{Logic program.}
    \label{fig:alg-sup-trad}
  \end{subfigure}
  \hfill
  \begin{subfigure}[b]{0.25\textwidth}
    \centering
    \begin{tikzpicture}[every text node part/.style={align=center}]
      \node[draw,double,minimum height=0.45cm] (x) at (0,0) {$x$};
      \node[draw,minimum height=0.45cm] (m) [right=0.5cm of x] {$M_\theta$};
      \node[draw,double,minimum height=0.45cm] (y) [right=0.5cm of m] {$y$};
      \draw[->] (x.east) -- (m.west);
      \draw[->] (m.east) -- (y.west);
      \draw[->, densely dashed] (y.south) -- ([yshift=-0.4cm]y.south) -- ([yshift=-0.4cm]m.south) node [midway, fill=white] {$\frac{\partial y}{\partial \theta}$} -- (m.south);
    \end{tikzpicture}
    \vspace{-0.07in}
    \caption{Neural model.}
    \label{fig:alg-sup-ml}
  \end{subfigure}
  \hfill
  \begin{subfigure}[b]{0.4\textwidth}
    \centering
    \begin{tikzpicture}
      \node[draw,double,minimum height=0.45cm] (x) at (0,0) {$x$};
      \node[draw,minimum height=0.45cm] (m) [right=0.5cm of x] {$M_\theta$};
      \node[draw,minimum height=0.45cm,densely dotted] (r) [right=0.5cm of m] {$r$};
      \node[draw,minimum height=0.45cm] (p) [right=0.5cm of r] {$P$};
      \node[draw,double,minimum height=0.45cm] (y) [right=0.5cm of p] {$y$};
      \draw[->] (x.east) -- (m.west);
      \draw[->] (m.east) -- (r.west);
      \draw[->] (r.east) -- (p.west);
      \draw[->] (p.east) -- (y.west);
      \draw[->, densely dashed] (y.south) -- ([yshift=-0.4cm]y.south) -- ([shift=({0.05cm,-0.4cm})]r.south) node [midway, fill=white] {$\frac{\partial y}{\partial r}$} -- ([xshift=0.05cm]r.south);
      \draw[->, densely dashed] ([xshift=-0.05cm]r.south) -- ([shift=({-0.05cm,-0.4cm})]r.south) -- ([yshift=-0.4cm]m.south) node [midway, fill=white] {$\frac{\partial r}{\partial \theta}$} -- (m.south);
    \end{tikzpicture}
    \vspace{-0.07in}
    \caption{A basic neurosymbolic program.}
    \label{fig:alg-sup-nesy}
  \end{subfigure}
  \caption{
    Comparison of different paradigms.
    Logic program $P$ accepts only structured input $r$ whereas neural model $M_\theta$ with parameter $\theta$ can operate on unstructured input $x$.
    Supervision is provided on data indicated in double boxes.
    Under {\em algorithmic supervision}, a neurosymbolic program must learn $\theta$ without supervision on $r$.
  }
  \label{fig:algorithmic-supervision}
  \Description{}
\end{figure}
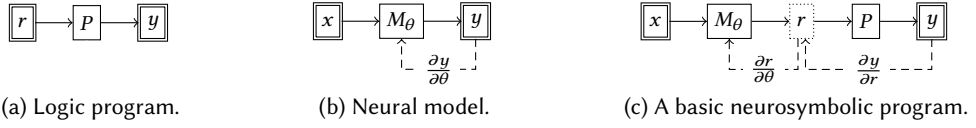

The two paradigms are complementary in nature.
For instance, a classical algorithm such as the logic program $P$ depicted in Figure \ref{fig:alg-sup-trad} is interpretable but operates on limited (e.g., structured) input $r$.
In contrast, a deep neural network such as $M_\theta$ depicted in
Figure \ref{fig:alg-sup-ml} can operate on rich (e.g., unstructured) input $x$ but is not interpretable.
Modern applications demand the capabilities of both paradigms.
Examples include question answering \cite{rajpurkar2016squad}, code completion \cite{chen2021evaluating}, and mathematical problem solving \cite{lewkowycz2022solving}, among many others.
For instance, code completion requires deep learning to comprehend programmer intent from the code context, and classical algorithms to ensure that the generated code is correct.
A natural and fundamental question then is how to program such applications by integrating the two paradigms.

Neurosymbolic programming is an emerging paradigm that aims to fulfill this goal \cite{chaudhuri2021neurosymbolic}.
It seeks to integrate symbolic knowledge and reasoning with neural architectures for better efficiency, interpretability, and generalizability than the neural or symbolic counterparts alone.
Consider the task of handwritten formula evaluation \cite{li2020closed}, which takes as input a formula as an image, and outputs a number corresponding to the result of evaluating it.
An input-output example for this task is $\langle x = \hwfsample{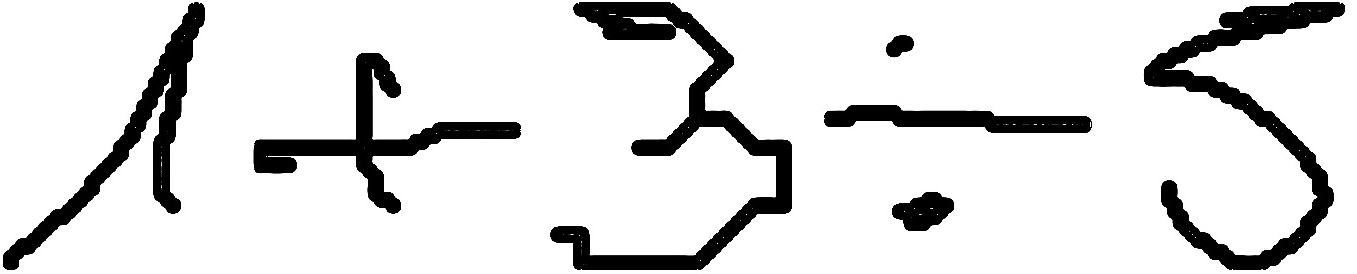}, y = 1.6 \rangle$.
A neurosymbolic program for this task, such as the one depicted in Figure~\ref{fig:alg-sup-nesy}, might first apply a convolutional neural network $M_\theta$ to the input image to obtain a structured intermediate form $r$ as a sequence of symbols [`1', `+', `3', `/', `5'], followed by a classical algorithm $P$ to parse the sequence, evaluate the parsed formula, and output the final result $1.6$.

Despite significant strides in individual neurosymbolic applications \cite{yi2018nsvqa, mao2019nscl, chen2020nerd, li2020closed, minervini2020ctp, wang2019satnet}, there is a lack of a language with compiler support to make the benefits of the neurosymbolic paradigm more widely accessible.
We set out to develop such a language and identified five key criteria that it should satisfy in order to be practical.
These criteria, annotated by the components of the neurosymbolic program in Figure \ref{fig:alg-sup-nesy}, are as follows:
\begin{enumerate}[leftmargin=0.7cm]
    \item A symbolic data representation for $r$ that supports diverse kinds of data, such as image, video, natural language text, tabular data, and their combinations.
    \item A symbolic reasoning language for $P$ that allows to express common reasoning patterns such as recursion, negation, and aggregation.
    \item An automatic and efficient differentiable reasoning engine for learning ($\frac{\partial{y}}{\partial{r}}$) under {\em algorithmic supervision}, i.e., supervision on observable input-output data $(x, y)$ but not $r$.
    \item The ability to tailor learning ($\frac{\partial{y}}{\partial{r}}$) to individual applications' characteristics, since non-continuous loss landscapes of logic programs hinder learning using a one-size-fits-all method.
    \item A mechanism to leverage and integrate with existing training pipelines ($\frac{\partial{r}}{\partial{\theta}}$), implementations of neural architectures and models $M_\theta$,
    and hardware (e.g., GPU) optimizations.
\end{enumerate}
In this paper, we present \ours, a language which satisfies the above criteria.
The key insight underlying \ours~ is its choice of three inter-dependent design decisions: a relational model for symbolic data representation, a declarative language for symbolic reasoning, and a provenance framework for differentiable reasoning.
We elaborate upon each of these decisions.

Relations can represent arbitrary graphs and are therefore well suited for representing symbolic data in \ours~ applications.
For instance, they can represent {\em scene graphs}, a canonical symbolic representation of images \cite{Johnson2015scenegraph}, or abstract syntax trees, a symbolic representation
of natural language text.
Further, they can be combined in a {\em relational database} to represent multi-modal data.
Relations are also a natural fit for probabilistic reasoning \cite{anton2015problog2}, which is necessary since symbols in $r$ produced by neural model $M_\theta$ can have associated probabilities.

Next, the symbolic reasoning program $P$ in \ours~ is specified in a declarative logic programming language.
The language extends Datalog \cite{abiteboul1995foundations}
and is expressive enough for programmers to specify complex domain knowledge patterns that the neural model $M_\theta$ would struggle with.
Datalog implementations can take advantage of optimizations from the literature on relational database systems.
This in turn enables efficient inference and learning since the logical domain knowledge specifications of the task at hand help reduce the burden of $M_\theta$, whose responsibilities are now less complex and more modular.
Finally, Datalog is rule-based, which makes programs easier to write, debug, and verify.
It also facilitates inferring them by leveraging techniques from program synthesis \cite{gulwani2017program} and ILP \cite{copper2022ilp}.

While symbolic reasoning offers many aforementioned benefits, it poses a fundamental challenge for learning parameter $\theta$.
Deep learning relies on gradient-based methods, enabled by the differentiable nature of the low-level activation functions that comprise $M_\theta$, to obtain $\frac{\partial{r}}{\partial{\theta}}$.
The key challenge then concerns how to support automatic and efficient differentiation of the high-level logic program $P$ to obtain $\frac{\partial{y}}{\partial{r}}$, which can be used in conjunction with $\frac{\partial{r}}{\partial{\theta}}$ to compute $\frac{\partial{y}}{\partial{\theta}}$.
\ours~ addresses this problem by leveraging the framework of {\em provenance semirings} \cite{provenancesemiring}.
The framework proposes a common algebraic structure for applications that define annotations (i.e., tags) for tuples and propagate the annotations from inputs to outputs of relational algebra (RA) queries.
One of our primary contributions is a novel adaptation of the framework for differentiable reasoning for an extended fragment of RA that includes recursion, negation, and aggregation.
\ours~ implements an extensible library of provenance structures including the extended max-min semiring and the top-$k$ proofs semiring \cite{huang2021scallop}.
We further demonstrate that different provenance structures enable different heuristics for the gradient calculations, providing an effective mechanism to tailor the learning process to individual applications' characteristics.

We have implemented a comprehensive and open-source toolchain for \ours~ in 45K LoC of Rust.
It includes a compiler, an interpreter, and PyTorch bindings to integrate \ours~ programs with existing machine learning pipelines.
We evaluate \ours~ using a suite of eight neurosymbolic applications that span the domains of image and video processing, natural language processing, planning, and knowledge graph querying, in a variety of learning settings such as supervised learning, reinforcement learning, rule learning, and contrastive learning.
Our evaluation demonstrates that \ours~ is expressive and yields solutions of comparable, and often times superior, accuracy than state-of-the-art models.
We show additional benefits of \ours's solutions in terms of runtime and data efficiency, interpretability, and generalizability.

Any programming language treatise would be remiss without acknowledging the language's lineage.
TensorLog \cite{cohen2017tensorlog} and DeepProbLog (DPL) \cite{manhaeve2021deepproblog} pioneered the idea of extending probabilistic logic programming languages (e.g., ProbLog \cite{anton2015problog2}) with differentiable reasoning.
Scallop was originally proposed in \cite{huang2021scallop} to improve the scalability of DPL by using Datalog instead of Prolog and relaxing its exact probabilistic semantics.
We build upon \cite{huang2021scallop} by extending its expressiveness, formalizing the semantics, developing a customizable provenance framework, and providing a practical toolchain.

The rest of the paper is organized as follows.
Section \ref{sec:overview} presents an illustrative overview of \ours.
Section \ref{sec:language} describes \ours's language for symbolic reasoning.
Section \ref{sec:reasoning-framework} presents the differentiable reasoning framework.
Section \ref{sec:implementation} describes our implementation of \ours.
Section \ref{sec:evaluation} empirically evaluates \ours~ on a benchmark suite.
Section \ref{sec:related} surveys related work and Section \ref{sec:conclusion} concludes.

\section{Illustrative Overview}
\label{sec:overview}

\begin{figure}
  \centering
  \footnotesize
  \begin{minipage}{0.4\textwidth}
  \begin{subfigure}[a]{\linewidth}
    \centering
    \footnotesize
    \begin{tabular}{ccc}
      \includegraphics[width=0.27\linewidth]{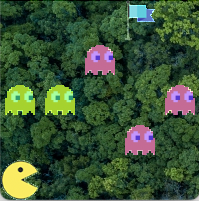}
      &
      \includegraphics[width=0.27\linewidth]{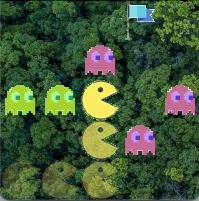}
      &
      \includegraphics[width=0.27\linewidth]{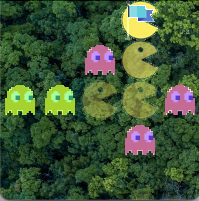}
      \\
      Step 0
      &
      Step 4
      &
      Step 7
    \end{tabular}
    \caption{Three states of one gameplay session.}
    \label{fig:pacman-maze-demo}
  \end{subfigure}
  
  \vspace{15px}
  
  \begin{subfigure}[b]{\linewidth}
    \centering
    \footnotesize
    \includegraphics[width=0.95\textwidth]{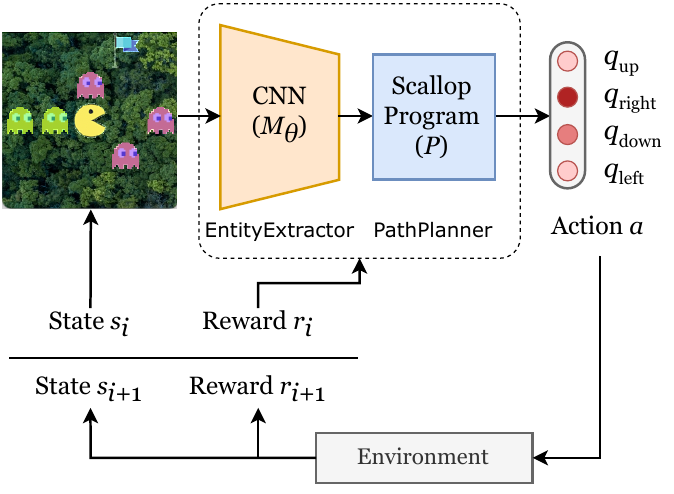}
    \caption{Architecture of application with \ours.}
    \label{fig:scallop-path-planner-flow}
  \end{subfigure}
  \end{minipage}
  \hfill
\begin{minipage}{0.55\textwidth}
\begin{subfigure}[c]{\linewidth}
\begin{lstlisting}[language=mypython,numbers=left,xleftmargin=0.1in]
class PacManAgent(torch.nn.Module):
 def __init__(self, grid_dim, cell_size):
  # initializations...
  self.extract_entities =
   EntityExtractor(grid_dim, cell_size)
  self.path_planner = ScallopModule(
   file="path_planner.scl",
   provenance="diff-top-k-proofs", k=1,
   input_mappings={"actor": cells,
    "goal": cells, "enemy": cells},
   output_mappings={"next_action": actions})

 def forward(self, game_state_image):
  actor, goal, enemy =
   self.extract_entities(game_state_image)
  next_action = self.path_planner(
   actor=actor, goal=goal, enemy=enemy)
  return next_action
\end{lstlisting}
\vspace{-5px}
\caption{Snippet of implementation in Python.}
\label{fig:scallop-path-planner-code}
\Description{}
\end{subfigure}
\end{minipage}
\caption{Illustration of a planning application PacMan-Maze in \ours.}
\label{fig:pacman-overall}
\end{figure}

We illustrate \ours~ using an reinforcement learning (RL) based planning application which we call PacMan-Maze.
The application, depicted in \figref{fig:pacman-maze-demo}, concerns an intelligent agent realizing a sequence of actions in a simplified version of the PacMan maze game.
The maze is an implicit $5 \times 5$ grid of cells.
Each cell is either empty or has an entity, which can be either the {\em actor} (PacMan),  the {\em goal} (flag), or an {\em enemy} (ghost).
At each step, the agent moves the actor in one of four directions: up, down, right, or left.
The game ends when the actor reaches the goal or hits an enemy.
The maze is provided to the agent as a raw image that is updated at each step, requiring the agent to process sensory inputs, extract relevant features, and logically plan the path to take.
Additionally, each session of the game has randomized initial positions of the actor, the goal, and the enemies.

Concretely, the game is modeled as a sequence of interactions between the agent and an environment, as depicted in \figref{fig:scallop-path-planner-flow}.
The game state $s_i \in S$ at step~$i$ is a $200 \times 200$ colored image ($S = \mathbb{R}^{200 \times 200 \times 3}$).
The agent proposes an action $a_i \in A = \{\inlinecodesmall{up}, \inlinecodesmall{down}, \inlinecodesmall{right}, \inlinecodesmall{left}\}$ to the environment, which generates a new image $s_{i + 1}$ as the next state.
The environment also returns a reward $r_i$ to the agent:
1 upon reaching the goal, and 0 otherwise.
This procedure repeats until the game ends or reaches a predefined limit on the number of steps.

A popular RL method to realize our application is $Q$-Learning. Its goal is to learn a function $Q : S \times A \rightarrow \mathbb{R}$ that returns the expected reward of taking action $a_i$ in state $s_i$.\footnote{We elide the Q-Learning algorithm as it is not needed to illustrate the neurosymbolic programming aspects of our example.}
Since the game states are images, we employ Deep $Q$-Learning \cite{mnih2015dqn}, which approximates the $Q$ function using a convolutional neural network (CNN) model with learned parameter~$\theta$.
An end-to-end deep learning based approach for our application involves training the model to predict the $Q$-value of each action for a given game state.
This approach takes 50K training episodes to achieve a 84.9\% test success rate, where a single episode is one gameplay session from start to end.

In contrast, a neurosymbolic solution using \ours~ only needs 50 training episodes to attain a 99.4\% test success rate.
\ours~ enables to realize these benefits of the neurosymbolic paradigm by decomposing the agent's task into separate neural and symbolic components, as shown in \figref{fig:scallop-path-planner-flow}.
These components perform sub-tasks that are ideally suited for their respective paradigms: the neural component perceives pixels of individual cells of the image at each step to identify the entities in them, while the symbolic component reasons about enemy-free paths from the actor to the goal to determine the optimal next action.
\figref{fig:scallop-path-planner-code} shows an outline of this architecture's implementation using the popular PyTorch framework.

Concretely, the neural component is still a CNN, but it now takes the pixels of a single cell in the input image at a time, and classifies the entity in it.
The implementation of the neural component (\inlinecodesmall{EntityExtractor}) is standard and elided for brevity.
It is invoked on lines 14-15 with input \inlinecodesmall{game\_state\_image}, a tensor in $\mathbb{R}^{200 \times 200 \times 3}$, and returns three $\mathbb{R}^{5 \times 5}$ tensors of entities.
For example, $\inlinecodesmall{actor}$ is an $\mathbb{R}^{5 \times 5}$ tensor and $\inlinecodesmall{actor}_{ij}$ is the probability of the actor being in cell $(i, j)$.
A representation of the entities is then passed to the symbolic component on lines 16-17, which derives the $Q$-value of each action.
The symbolic component, which is configured on lines 6-11, comprises the \ours~ program shown in \figref{fig:scallop-q-network-scallop}.
We next illustrate the three key design decisions of \ours~ outlined in Section \ref{sec:introduction} with respect to this program.

{\bf {\em Relational Model.}}
In \ours, the primary data structure for representing symbols is a {\em relation}.
In our example, the game state can be symbolically described by the kinds of entities that occur in the discrete cells of a $5 \times 5$ grid.
We can therefore represent the input to the symbolic component using binary relations for the three kinds of entities: \inlinecodesmall{actor}, \inlinecodesmall{goal}, and \inlinecodesmall{enemy}.
For instance, the fact \inlinecodesmall{actor(2,3)} indicates that the actor is in cell (2,3).
Likewise, since there are four possible actions, the output of the symbolic component is represented by a unary relation \inlinecodesmall{next\_action}.

\begin{figure}
\begin{lstlisting}[language=scallop,numbers=left,firstnumber=1,xleftmargin=.05\textwidth]
// File path_planner.scl
type actor(x: i32, y: i32), goal(x: i32, y: i32), enemy(x: i32, y: i32)

const UP = 0, DOWN = 1, RIGHT = 2, LEFT = 3
rel safe_cell(x, y) = range(0, 5, x), range(0, 5, y), not enemy(x, y)
rel edge(x, y, x, yp, UP) = safe_cell(x, y), safe_cell(x, yp), yp == y + 1
// Rules for DOWN, RIGHT, and LEFT edges are omitted...

rel next_pos(p, q, a) = actor(x, y), edge(x, y, p, q, a)
rel path(x, y, x, y) = next_pos(x, y, _)
rel path(x1, y1, x3, y3) = path(x1, y1, x2, y2), edge(x2, y2, x3, y3, _)
rel next_action(a) = next_pos(p, q, a), path(p, q, r, s), goal(r, s)
\end{lstlisting}
\caption{The logic program of the PacMan-Maze application in \ours.}
\label{fig:scallop-q-network-scallop}
\Description{}
\end{figure}

Symbols extracted from unstructured inputs by neural networks have associated probabilities, such as the $\mathbb{R}^{5 \times 5}$ tensor \inlinecodesmall{actor} produced by the neural component in our example (line 14 of \figref{fig:scallop-path-planner-code}).
\ours~ therefore allows to associate tuples with probabilities, e.g. $0.96::\inlinecodesmall{actor(2,3)}$, to indicate that the actor is in cell (2,3) with probability 0.96.
More generally, \ours~ enables the conversion of tensors in the neural component to and from relations in the symbolic component via input-output mappings (lines 9-11 in \figref{fig:scallop-path-planner-code}), allowing the two components to exchange information seamlessly.


{\bf {\em Declarative Language.}}
Another key consideration in a neurosymbolic language concerns what constructs to provide for symbolic reasoning.
\ours~ uses a declarative language based on Datalog, which we present in Section \ref{sec:language} and illustrate here using the program in \figref{fig:scallop-q-network-scallop}.
The program realizes the symbolic component of our example using a set of logic rules.
Following Datalog syntax, they are ``if-then'' rules, read right to left, with commas denoting conjunction.

Recall that we wish to determine an action $a$ (up, down, right, or left) to a cell $(p, q)$
that is adjacent to the actor's cell $(x, y)$ such that there is an enemy-free path from $(p, q)$ to the goal's cell $(r, s)$.
The nine depicted rules succinctly compute this sophisticated reasoning pattern by building successively complex relations, with the final rule (on line 14) computing all such actions.\footnote{
We elide showing an auxiliary relation of all grid cells tagged with probability 0.99 which serves as the penalty for taking a step.
Thus, longer paths are penalized more, driving the symbolic program to prioritize moving closer to the goal.
}

The arguably most complex concept is the \inlinecodesmall{path} relation which is recursively defined (on lines 10-11).
Recursion allows to define the pattern succinctly, enables the trained application to generalize to grids arbitrarily larger than $5 \times 5$ unlike the purely neural version, and makes the pattern more amenable to synthesis from input-output examples.
Besides recursion, \ours~ also supports negation and aggregation; together, these features render the language adequate for specifying common high-level reasoning patterns in practice.

{\bf {\em Differentiable Reasoning.}}
With the neural and symbolic components defined, the last major consideration concerns how to train the neural component using only end-to-end supervision.
In our example, supervision is provided in the form of a reward of 1 or 0 per gameplay session, depending upon whether or not the sequence of actions by the agent successfully led the actor to the goal without hitting any enemy.
This form of supervision, called algorithmic or weak supervision, alleviates the need to label intermediate relations at the interface of the neural and symbolic components, such as the \inlinecodesmall{actor}, \inlinecodesmall{goal}, and \inlinecodesmall{enemy} relations.
However, this also makes it challenging to learn the gradients for the tensors of these relations, which in turn are needed to train the neural component using gradient-descent techniques.

The key insight in \ours~ is to exploit the structure of the logic program to guide the gradient calculations.
The best heuristic for such calculations depends on several application characteristics such as the amount of available data, reasoning patterns, and the learning setup.
\ours~ provides a convenient interface for the user to select from a library of built-in heuristics.
Furthermore, since all of these heuristics follow the structure of the logic program,
\ours~ implements them uniformly as instances of a general and extensible \textit{provenance framework}, described in \secref{sec:reasoning-framework}.
For our example, line 8 in \figref{fig:scallop-path-planner-code} specifies \inlinecodesmall{diff-top-k-proofs} with \inlinecodesmall{k=1} as the heuristic to use, which is the default in \ours~ that works best for many applications, as we demonstrate in \secref{sec:evaluation}.

\section{Language}
\label{sec:language}

We provide an overview of \ours's language for symbolic reasoning which we previously illustrated in the program shown in \figref{fig:scallop-q-network-scallop}.
Appendix \ref{app:scallop-surface-language} provides the formal syntax of the language.
Here, we illustrate each of the key constructs using examples of inferring kinship relations.

\subsection{Data Types}
\label{sec:language-data-types}

The fundamental data type in \ours~ is set-valued relations comprising tuples of statically-typed primitive values.
The primitive data types include signed and unsigned integers of various sizes (e.g. \inlinecode{i32}, \inlinecode{usize}), single- and double-precision floating point numbers (\inlinecode{f32}, \inlinecode{f64}), boolean (\inlinecode{bool}), character (\inlinecode{char}), and string (\inlinecode{String}).
The following example declares two binary relations, \inlinecode{mother} and \inlinecode{father}:
\begin{lstlisting}[language=scallop,numbers=none,xleftmargin=.05\textwidth]
type mother(c: String, m: String), father(c: String, f: String)
\end{lstlisting}
Values of relations can be specified via individual tuples or a set of tuples of constant literals:
\begin{lstlisting}[language=scallop,numbers=none,xleftmargin=.05\textwidth]
rel mother("Bob", "Christine")                   // Christine is Bob's mother
rel father = {("Alice", "Bob"), ("John", "Bob")} // Bob is father of two kids
\end{lstlisting}
As a shorthand, primitive values can be named and used as constant variables:
\begin{lstlisting}[language=scallop,numbers=none,xleftmargin=.05\textwidth]
const FATHER = 0, MOTHER = 1, GRANDMOTHER = 2, ... // other relationships
rel composition(FATHER, MOTHER, GRANDMOTHER) // father's mother is grandmother
\end{lstlisting}
Type declarations are optional since \ours~ supports type inference.
The type of the \inlinecode{composition} relation is inferred as \inlinecode{(usize, usize, usize)} since the default type of unsigned integers is \inlinecode{usize}.

\subsection{(Horn) Rules}
\label{sec:language-horn-rules}

Since \ours's language is based on Datalog, it supports ``if-then'' rule-like Horn clauses.
Each rule is composed of a head atom and a body, connected by the symbol \inlinecode{:-} or \inlinecode{=}.
The following code shows two rules defining the \inlinecode{grandmother} relation.
Conjunction is specified using ``\inlinecode{,}''-separated atoms within the rule body whereas disjunction is specified by multiple rules with the same head predicate.
Each variable appearing in the head must also appear in some positive atom in the body (we introduce negative atoms below).

\begin{lstlisting}[language=scallop,numbers=none,xleftmargin=.05\textwidth]
rel grandmother(a, c) :- father(a, b), mother(b, c) // father's mother
rel grandmother(a, c) :- mother(a, b), mother(b, c) // mother's mother
\end{lstlisting}
Conjunctions and disjunctions can also be expressed using logical connectives like \inlinecode{and}, \inlinecode{or}, and \inlinecode{implies}.
For instance, the following rule is equivalent to the above two rules combined.
\begin{lstlisting}[language=scallop,numbers=none,xleftmargin=.05\textwidth]
rel grandmother(a, c) = (mother(a, b) or father(a, b)) and mother(b, c)
\end{lstlisting}
\ours~ supports value creation by means of foreign functions (FFs).
FFs are polymorphic and include arithmetic operators such as \inlinecode{+} and \inlinecode{-}, comparison operators such as \inlinecode{!=} and \inlinecode{>=}, type conversions such as \inlinecode{[i32] as String}, and built-in functions like \inlinecode{\$hash} and \inlinecode{\$string\_concat}.
They only operate on primitive values but not relational tuples or atoms.
The following example shows how strings are concatenated together using FF, producing the result \inlinecode{full\_name("Alice Lee")}.
\begin{lstlisting}[language=scallop,numbers=none,xleftmargin=.05\textwidth]
rel first_name("Alice"), last_name("Lee")
rel full_name($string_concat(x, " ", y)) = first_name(x), last_name(y)
\end{lstlisting}
Note that FFs can fail due to runtime errors such as division-by-zero and integer overflow, in which case the computation for that single fact is omitted.
In the example below, when dividing 6 by \inlinecode{denominator}, the result is not computed for denominator \inlinecode{0} since it causes a FF failure:
\begin{lstlisting}[language=scallop,numbers=none,xleftmargin=.05\textwidth]
rel denominator = {0, 1, 2}
rel result(6 / x) = denominator(x) // result contains only integers 3 and 6
\end{lstlisting}
The purpose of this semantics is to support probabilistic extensions (Section \ref{sec:language-probabilistic-extensions}) rather than silent suppression of runtime errors.
When dealing with floating-point numbers, tuples with \texttt{NaN} (not-a-number) are also discarded.

\textit{Recursion.}
A relation $r$ is dependent on $s$ if an atom $s$ appears in the body of a rule with head atom $r$.
A \textit{recursive} relation is one that depends on itself, directly or transitively.
The following rule derives additional \inlinecode{kinship} facts by composing existing \inlinecode{kinship} facts using the \inlinecode{composition} relation.
\begin{lstlisting}[language=scallop,numbers=none,xleftmargin=.05\textwidth]
rel kinship(r3,a,c) = kinship(r1,a,b), kinship(r2,b,c), composition(r1,r2,r3)
\end{lstlisting}

\textit{Negation.}
\ours~ supports stratified negation using the \inlinecode{not} operator on atoms in the rule body.
The following example shows a rule defining the \inlinecode{has\_no\_children} relation as any person \inlinecode{p} who is neither a father nor a mother.
Note that we need to bound \inlinecode{p} by a positive atom \inlinecode{person} in order for the rule to be well-formed.
\begin{lstlisting}[language=scallop,numbers=none,xleftmargin=.05\textwidth]
rel person = {"Alice", "Bob", "Christine"} // can omit () since arity is 1
rel has_no_children(p) = person(p) and not father(_, p) and not mother(_, p)
\end{lstlisting}
A relation $r$ is \textit{negatively} dependent on $s$ if a negated atom $s$ appears in the body of a rule with head atom $r$.
In the above example, \inlinecode{has\_no\_children} negatively depends on \inlinecode{father}.
A relation cannot be negatively dependent on itself, directly or transitively, as \ours~ supports only stratified negation.
The following rule is rejected by the compiler, as the negation is not stratified:
\begin{lstlisting}[language=scallop,numbers=none,xleftmargin=.05\textwidth]
rel something_is_true() = not something_is_true() // compilation error!
\end{lstlisting}

\textit{Aggregation.}
\ours~ also supports stratified aggregation.
The set of built-in aggregators include common ones such as \inlinecode{count}, \inlinecode{sum}, \inlinecode{max}, and first-order quantifiers \inlinecode{forall} and \inlinecode{exists}.
Besides the operator, the aggregation specifies the binding variables, the aggregation body to bound those variables, and the result variable(s) to assign the result.
The aggregation in the example below reads,
``variable \inlinecode{n} is assigned the count of \inlinecode{p}, such that \inlinecode{p} is a person'':
\begin{lstlisting}[language=scallop,numbers=none,xleftmargin=.05\textwidth]
rel num_people(n) = n := count(p: person(p))
\end{lstlisting}
In the rule, \inlinecode{p} is the binding variable and \inlinecode{n} is the result variable.
Depending on the aggregator, there could be multiple binding variables or multiple result variables.
Further, \ours~ supports SQL-style group-by using a \inlinecode{where} clause in the aggregation.
In the following example, we compute the number of children of each person \inlinecode{p}, which serves as the group-by variable.
\begin{lstlisting}[language=scallop,numbers=none,xleftmargin=.05\textwidth]
rel parent(a, b) = father(a, b) or mother(a, b)
rel num_child(p, n) = n := count(c: parent(c, p) where p: person(p))
\end{lstlisting}
Finally, quantifier aggregators such as \inlinecode{forall} and \inlinecode{exists} return one boolean variable.
For instance, in the aggregation below, variable \inlinecode{sat} is assigned the truthfulness (\inlinecode{true} or \inlinecode{false}) of the following statement: ``for all \inlinecode{a} and \inlinecode{b}, if \inlinecode{b} is \inlinecode{a}'s father, then \inlinecode{a} is \inlinecode{b}'s son or daughter''.
\begin{lstlisting}[language=scallop,numbers=none,xleftmargin=.05\textwidth]
rel integrity_constraint(sat) =
    sat := forall(a, b: father(a, b) implies (son(b, a) or daughter(b, a)))
\end{lstlisting}
There are a couple of syntactic checks on aggregations.
First, similar to negation, aggregation also needs to be stratified---a relation cannot be dependent on itself through an aggregation.
Second, the binding variables must be bounded by a positive atom in the body of the aggregation.

\subsection{Probabilistic Extensions}
\label{sec:language-probabilistic-extensions}

Although \ours~ is designed primarily for neurosymbolic programming, its syntax also supports probabilistic programming.
This is especially useful when debugging \ours~ code before integrating it with a neural network.
Consider a machine learning programmer who wishes to extract structured relations from a natural language sentence ``Bob takes his daughter Alice to the beach''.
The programmer could imitate a neural network producing a probability distribution of kinship relations between Alice (\inlinecodesmall{A}) and Bob (\inlinecodesmall{B}) as follows:
\begin{lstlisting}[language=scallop,numbers=none,xleftmargin=.05\textwidth]
rel kinship = {0.95::(FATHER, A, B); 0.01::(MOTHER, A, B); ... }
\end{lstlisting}
Here, we list out all possible kinship relations between Alice and Bob.
For each of them, we use the syntax \inlinecode{[PROB]::[TUPLE]} to tag the kinship tuples with probabilities.
The semicolon ``\inlinecode{;}'' separating them specifies that they are mutually exclusive---Bob cannot be both the mother and father of Alice.

\ours~ also supports operators to sample from probability distributions. They share the same surface syntax as aggregations, allowing sampling with group-by.
The following rule deterministically picks the most likely kinship relation between a given pair of people \inlinecode{a} and \inlinecode{b}, which are implicit group-by variables in this aggregation.
As the end, only one fact, \inlinecode{0.95::top\_1\_kinship(FATHER, A, B)}, will be derived according to the above probabilities.
\begin{lstlisting}[language=scallop,numbers=none,xleftmargin=.05\textwidth]
rel top_1_kinship(r, a, b) = r := top<1>(rp: kinship(rp, a, b))
\end{lstlisting}

Other types of sampling are also supported, including categorical sampling (\inlinecode{categorical<K>}) and uniform sampling (\inlinecode{uniform<K>}), where a static constant \inlinecode{K} denotes the number of trials.

Finally, rules can also be tagged by probabilities which can reflect their confidence.
The following rule states that a grandmother's daughter is one's mother with 90\% confidence.
\begin{lstlisting}[language=scallop,numbers=none,xleftmargin=.05\textwidth]
rel 0.9::mother(a, c) = grandmother(a, b) and daughter(b, c)
\end{lstlisting}
Probabilistic rules are syntactic sugar.
They are implemented by introducing in the rule's body an auxiliary 0-ary (i.e., boolean) fact that is regarded true with the tagged probability.

\section{Reasoning Framework}
\label{sec:reasoning-framework}

\begin{figure}
  \footnotesize
  \begin{minipage}{0.44\textwidth}
    \footnotesize
    \[
      \begin{array}{rccl}
      \text{(Tag)} & t & \in & T \\
      \text{(False)} & \mathbf{0} & \in & T \\
      \text{(True)} & \mathbf{1} & \in & T \\
      \text{(Disjunction)} & \oplus & : & T \times T \rightarrow T \\
      \text{(Conjunction)} & \otimes & : & T \times T \rightarrow T \\
      \text{(Negation)} & \ominus & : & T \rightarrow T \\
      \text{(Saturation)} & \oeq & : & T \times T
      \rightarrow \text{Bool} \\
      \end{array}
    \]
    \vspace{-0.15in}
    \captionof{figure}{Algebraic interface for provenance.}
    \label{fig:core-provenance-interface}
  \end{minipage}
  \hfill
  \begin{minipage}{0.52\textwidth}
    \footnotesize
    \[
      \begin{array}{rcrl}
      \text{(Predicate)} & p \\
      \text{(Aggregator)} & g & ::= & \inlinecode{count} \sep \inlinecode{sum} \sep \inlinecode{max} \sep \inlinecode{exists} \sep \dots \\
      \text{(Expression)} & e & ::= & p \sep \gamma_g(e) \sep \pi_\alpha(e) \sep \sigma_\beta(e) \\
                          &   & \sep & e_1 \cup e_2 \sep e_1 \times e_2 \sep e_1 - e_2 \\
      \text{(Rule)} & r & ::= & p \leftarrow e \\
      \text{(Stratum)} & s & ::= & \{ r_1, \dots, r_n \} \\
      \text{(Program)} & \overline{s} & ::= & s_1; \dots; s_n \\
      \end{array}
    \]
    \vspace{-0.15in}
    \captionof{figure}{Abstract syntax of core fragment of \ram.}
    \label{fig:core-sclram-syntax}
  \end{minipage}
  \Description{}
  \vspace{-0.1in}
\end{figure}

The preceding section presented \ours's surface language for use by programmers to express discrete reasoning.
However, the language must also support differentiable reasoning to enable end-to-end
training.
In this section, we formally define the semantics of the language by means of a provenance framework.
We show how \ours~ uniformly supports different reasoning modes---discrete, probabilistic, and differentiable---simply by defining different provenances.

We start by presenting the basics of our provenance framework (\secref{sec:provenance-framework}).
We then present a low-level representation \ram, its operational semantics, and its interface to the rest of a \ours~ application (Sections \ref{sec:core-syntax}-\ref{sec:prov-extensions}).
We next present differentiation and three different provenances for differentiable reasoning
(\secref{sec:prov-structures}).
Lastly, we discuss practical considerations (\secref{sec:prov-discussions}).

\subsection{Provenance Framework}
\label{sec:provenance-framework}

\ours's provenance framework enables to tag and propagate additional information alongside relational tuples in the logic program's execution.
The framework is based on the theory of \textit{provenance semirings} \cite{provenancesemiring}.
\figref{fig:core-provenance-interface} defines \ours's algebraic interface for provenance.
We call the additional information a \textit{tag} $t$ from a \textit{tag space} $T$.
There are two distinguished tags, $\mathbf{0}$ and $\mathbf{1}$, representing unconditionally \textit{false} and \textit{true} tags.
Tags are propagated through operations of binary \textit{disjunction} $\oplus$, binary \textit{conjunction} $\otimes$, and unary \textit{negation} $\ominus$ resembling logical \textit{or}, \textit{and}, and \textit{not}.
Lastly, a \textit{saturation} check $\oeq$ serves as a customizable stopping mechanism for fixed-point iteration.

All of the above components combined form a 7-tuple $(T, \mathbf{0}, \mathbf{1}, \oplus, \otimes, \ominus, \oeq)$ which we call a \textit{provenance} $T$.
\ours~ provides a built-in library of provenances and users can add custom provenances simply by implementing this interface.

\vspace{-0.05in}
\begin{example}
  \inlinecodesmall{max-min-prob}
  ($\inlinecodesmall{mmp}) \triangleq ([0, 1], 0, 1, \text{max}, \text{min}, \lambda x . (1 - x), \texttt{==})$,
  is a built-in \textit{probabilistic provenance}, where tags are numbers between 0 and 1 that are propagated with max and min.
  The tags do not represent true probabilities but are merely an approximation.
  We discuss richer provenances for more accurate probability calculations later in this section.
  \label{example:max-min-prob}
\end{example}

A provenance must satisfy a few properties. 
First, the 5-tuple $(T, \mathbf{0}, \mathbf{1}, \oplus, \otimes)$ should form a semiring.
That is, $\mathbf{0}$ is the additive identity and annihilates under multiplication, $\mathbf{1}$ is the multiplicative identity, $\oplus$ and $\otimes$ are associative and commutative, and $\otimes$ distributes over $\oplus$.
To guarantee the existence of fixed points (which are discussed in \secref{sec:core-semantics}), it must also be \textit{absorptive}, i.e., $t_1 \oplus (t_1 \otimes t_2) = t_1$ \cite{semiring-provenance-for-fixed-point-logic}.
Moreover, we need $\ominus~ \mathbf{0} = \mathbf{1}$, $\ominus~ \mathbf{1} = \mathbf{0}$, $\mathbf{0} \oneq \mathbf{1}$, $\mathbf{0} \oeq \mathbf{0}$, and $\mathbf{1} \oeq \mathbf{1}$.
A provenance which violates an individual property (e.g. absorptive) is still useful to applications that do not use the affected features (e.g. recursion) or if the user simply wishes to bypass the restrictions.


\subsection{\ram~ Intermediate Language}
\label{sec:core-syntax}

\begin{figure}
  \footnotesize
  \[
    \begin{array}{rlcccl}
      \text{(Constant)} & \mathbb{C} & \ni & c & ::= & int \sep bool \sep str \sep \dots \\
      \text{(Tuple)} & \mathbb{U} & \ni & u & ::= & c \sep (u_1, \dots, u_n)\\
      \text{(Tagged-Tuple)} & \mathbb{U}_T & \ni & u_t & ::= & t :: u \\
      \text{(Fact)} & \mathbb{F} & \ni & f & ::= & p(u) \\
      \text{(Tagged-Fact)} & \mathbb{F}_T & \ni & f_t & ::= & t :: p(u) \\
    \end{array}
    \begin{array}{rcclcl}
      \text{(Tuples)} & U & \in & \mathcal{U} & \triangleq & \mathcal{P}(\mathbb{U}) \\
      \text{(Tagged-Tuples)} & U_T & \in & \mathcal{U}_T & \triangleq & \mathcal{P}(\mathbb{U}_T) \\
      \text{(Facts)} & F & \in & \mathcal{F} & \triangleq & \mathcal{P}(\mathbb{F}) \\
      \text{(Database)} & F_T & \in & \mathcal{F}_T & \triangleq & \mathcal{P}(\mathbb{F}_T) \\
    \end{array}
  \]
  \caption{Semantic domains for \ram.}
  \label{fig:sclram-operation-domain}
  \Description{}
    \vspace{-0.1in}
\end{figure}

\ours~ programs are compiled to a low-level representation called \ram.
\figref{fig:core-sclram-syntax} shows the abstract syntax of a core fragment of \ram.
Expressions resemble queries in an extended relational algebra. They operate over relational predicates ($p$) using unary operations for
aggregation ($\gamma_g$ with aggregator $g$),
projection ($\pi_\alpha$ with mapping $\alpha$), and
selection ($\sigma_\beta$ with condition $\beta$), and binary operations
union ($\cup$), product ($\times$), and difference ($-$).

A rule $r$ in \ram~ is denoted $p \leftarrow e$, where predicate $p$ is the rule head and expression $e$ is the rule body.
An unordered set of rules combined form a stratum $s$, and a sequence of strata $s_1; \dots; s_n$ constitutes a \ram~ program.
Rules in the same stratum have distinct head predicates.
Denoting the set of head predicates in stratum $s$ by $P_{s}$, we also require $P_{s_i} \cap P_{s_j} = \emptyset$ forall $i \neq j$ in a program.
Stratified negation and aggregation from the surface language is enforced as syntax restrictions in \ram:
within a rule in stratum $s_i$, if a relational predicate $p$ is used under aggregation ($\gamma$) or right-hand-side of difference ($-$),
that predicate $p$ cannot appear in $P_{s_j}$ if $j \geq i$.

We next define the semantic domains in \figref{fig:sclram-operation-domain} which are used subsequently to define the semantics of \ram.
A tuple $u$ is either a constant or a sequence of tuples.
A fact $p(u) \in \mathbb{F}$ is a tuple $u$ named under a relational predicate $p$.
Tuples and facts can be tagged, forming \textit{tagged tuples} ($t :: u$) and \textit{tagged facts} ($t :: p(u)$).
Given a set of tagged tuples $U_T$, we say $U_T \contains u$ iff there exists a $t$ such that $t :: u \in U_T$.
A set of tagged facts form a database $F_T$.
We use bracket notation $F_T[p]$ to denote the set of tagged facts in $F_T$ under predicate $p$.

\vspace{-0.05in}
\subsection{Operational Semantics of \ram}
\label{sec:core-semantics}

We now present the operational semantics for our core fragment of \ram~in Fig. \ref{fig:sclram-semantics}.
A \ram~ program $\overline{s}$ takes as input an \textit{extensional database} (EDB) $F_T$, and returns an \textit{intentional database} (IDB) $F_T' = \sem{\overline{s}}(F_T)$.
The semantics is conditioned on the underlying provenance $T$.
We call this \textit{tagged semantics}, as opposed to the \textit{untagged semantics} found in traditional Datalog.

\textit{Basic Relational Algebra.}
Evaluating an expression in \ram~ yields a set of tagged tuples according to the rules defined at the top of \figref{fig:sclram-semantics}.
A predicate $p$ evaluates to all facts under that predicate in the database.
Selection filters tuples that satisfy condition $\beta$, and projection transforms tuples according to mapping $\alpha$.
The mapping function $\alpha$ is partial: it may fail since it can apply foreign functions to values.
A tuple in a union $e_1 \cup e_2$ can come from either $e_1$ or $e_2$.
In (cartesian) product, each pair of incoming tuples combine and we use $\otimes$ to compute the tag conjunction.

\begin{figure}
  \footnotesize
  \textbf{Expression semantics}
  \hfill
  \framebox[1.1\width]{$
    \alpha : \mathbb{U} \rightharpoonup \mathbb{U}, \quad
    \beta : \mathbb{U} \rightarrow \text{Bool},\quad
    g : \mathcal{U} \rightarrow \mathcal{U}, \quad
    \sem{e} : \mathcal{F}_T \rightarrow \mathcal{U}_T
  $}
  $$
  \inferrule*[right=\anno{Predicate}]
    {t :: p(u) \in F_T}
    {t :: u \in \sem{p}(F_T)}
  \hspace*{5pt}
  \inferrule*[right=\anno{Select}]
    {t :: u \in \sem{e}(F_T) \\ \beta(u) = \text{true}}
    {t :: u \in \sem{\sigma_\beta(e)}(F_T)}
  \hspace*{5pt}
  \inferrule*[right=\anno{Project}]
    {t :: u \in \sem{e}(F_T) \\ u' = \alpha(u)}
    {t :: u' \in \sem{\pi_\alpha(e)}(F_T)}
  $$
  $$
  \inferrule*[right=\anno{Union}]
    {t :: u \in \sem{e_1}(F_T) \cup \sem{e_2}(F_T)}
    {t :: u \in \sem{e_1 \cup e_2}(F_T)}
  \quad
  \inferrule*[right=\anno{Product}]
    {t_1 :: u_1 \in \sem{e_1}(F_T) \\ t_2 :: u_2 \in \sem{e_2}(F_T)}
    {(t_1 \otimes t_2) :: (u_1, u_2) \in \sem{e_1 \times e_2}(F_T)}
  $$
  $$
  \inferrule*[right=\anno{Diff-1}]
    {t :: u \in \sem{e_1}(F_T) \\ \sem{e_2}(F_T) \nvDash u}
    {t :: u \in \sem{e_1 - e_2}(F_T)}
  \quad
  \inferrule*[right=\anno{Diff-2}]
    {t_1 :: u \in \sem{e_1}(F_T) \\ t_2 :: u \in \sem{e_2}(F_T)}
    {(t_1 \otimes (\ominus~ t_2)) :: u \in \sem{e_1 - e_2}(F_T)}
  $$
  $$
  \inferrule*[right=\anno{Aggregate}]
    {
      X_T \subseteq \sem{e}(F_T) \\
      \{ t_i :: u_i \}_{i=1}^n = X_T \\
      \{ \overline{t}_j :: \overline{u}_j \}_{j=1}^m = \sem{e}(F_T) - X_T \\
      u \in g(\{u_i\}_{i=1}^n)
    }
    {\textstyle (\bigotimes_{i=1}^n t_i) \otimes (\bigotimes_{j=1}^m (\ominus~ \overline{t}_j)) :: u \in \sem{\gamma_g(e)}(F_T)}
  $$
  \textbf{Rule semantics}
  \hfill
  \framebox[1.1\width]{$
    \left\langle . \right\rangle : \mathcal{U}_T \rightarrow \mathcal{U}_T, \quad
    \sem{r} : \mathcal{F}_T \rightarrow \mathcal{F}_T$}
  $$
  \textsc{(Normalize)} \quad
  \left\langle U_T \right\rangle = \{ (\textstyle \bigoplus_{i=1}^n t_i) :: u \sep t_1 :: u, \dots, t_n :: u ~\text{are all tagged-tuples in}~ U_T ~\text{with the same tuple}~ u \}
  $$
  $$
  \inferrule*[right=\anno{Rule-1}]
    {t^\old :: u \in \sem{p}(F_T) \\ \left\langle \sem{e}(F_T) \right\rangle \nvDash u}
    {t^\old :: p(u) \in \sem{p \leftarrow e}(F_T)}
  \quad
  \inferrule*[right=\anno{Rule-2}]
    {t^\new :: u \in \left\langle \sem{e}(F_T) \right\rangle \\ \sem{p}(F_T) \nvDash u}
    {t^\new :: p(u) \in \sem{p \leftarrow e}(F_T)}
  $$
  $$
  \inferrule*[right=\anno{Rule-3}]
    {t^\old :: u \in \sem{p}(F_T) \\ t^\new :: u \in \left\langle \sem{e}(F_T) \right\rangle}
    {(t^\old \oplus t^\new) :: p(u) \in \sem{p \leftarrow e}(F_T)}
  $$
  \textbf{Stratum and Program semantics}
  \hfill
  \framebox[1.1\width]{$\mathbf{lfp}^\circ : (\mathcal{F}_T \rightarrow \mathcal{F}_T) \rightarrow (\mathcal{F}_T \rightarrow \mathcal{F}_T), \quad \sem{s}, \sem{\overline{s}} : \mathcal{F}_T \rightarrow \mathcal{F}_T$}
  \begin{align*}
    \textsc{(Saturation)} &\quad
    F_T^\old \circeq F_T^\new ~\text{iff}~ \forall t^\new :: p(u) \in F_T^\new, \exists t^\old :: p(u) \in F_T^\old ~\text{such that}~ t^\old \oeq t^\new
    \\
    \textsc{(Fixpoint)} &\quad
    \mathbf{lfp}^\circ (h) = h \circ \dots \circ h = h^n ~\text{if there exists a minimum}~ n > 0,~\text{such that}~ h^{n}(F_T) \circeq h^{n+1}(F_T)
    \\
    \textsc{(Stratum)} &\quad
    \sem{s} = \mathbf{lfp}^\circ(\lambda F_T . (F_T - \textstyle \bigcup_{p \in P_s} F_T[p]) \cup (\textstyle \bigcup_{r \in s} \sem{r}(F_T)))
    \\
    \textsc{(Program)} &\quad
    \sem{\overline{s}} = \sem{s_n} \circ \dots \circ \sem{s_1}, ~\text{where}~ \overline{s} = s_1; \dots; s_n.
  \end{align*}
  \vspace{-0.16in}
  \caption{Operational semantics of core fragment of \ram.}
  \label{fig:sclram-semantics}
  \Description{}
  \vspace{-0.16in}
\end{figure}

\begin{figure}
  \centering
  \footnotesize
  \begin{subfigure}{\textwidth}
    \centering
    \framebox[1.05\width]{\inlinescl{rel safe\_cell(x, y) = grid\_cell(x, y), not enemy(x, y)}}
    \quad
    $\rightarrow$
    \quad
    \framebox[1.1\width]{$\inlinecode{safe\_cell} \leftarrow \inlinecode{grid\_cell} - \inlinecode{enemy}$}
  \end{subfigure}

  \vspace{5px}

  \begin{subfigure}{0.30\textwidth}
    \centering
    \begin{tikzpicture}
      \node[draw] (g) {$\sem{\inlinecode{g}}(F)$};
      \node[right=0.5cm of g] (gp) {};
      \node[draw,right=0.5cm of gp] (e) {$\sem{\inlinecode{e}}(F)$};
      \node[draw,below=0.4cm of gp] (s) {$\sem{\inlinecode{g} - \inlinecode{e}}(F)$};

      \node[rectangle,above=0.0cm of g] (gcontent) {\begin{tabular}{r}$(1, 2)$ \\ $(2, 3)$\end{tabular}};
      \node[rectangle,above=0.0cm of e] (econtent) {\begin{tabular}{r}$(2, 3)$\end{tabular}};
      \node[rectangle,below=0.0cm of s] (scontent) {\begin{tabular}{r}$(1, 2)$ \\ {} \end{tabular}};

      \draw[->] (g) -- (s.north west);
      \draw[->] (e) -- (s.north east);
    \end{tikzpicture}
    \caption{Untagged semantics}
    \label{fig:negation-example-datalog}
  \end{subfigure}
  \hfill
  \begin{subfigure}{0.30\textwidth}
    \centering
    \begin{tikzpicture}
      \node[draw] (g) {$\sem{\inlinecode{g}}(F_T)$};
      \node[right=0.5cm of g] (gp) {};
      \node[draw,right=0.5cm of gp] (e) {$\sem{\inlinecode{e}}(F_T)$};
      \node[draw,below=0.4cm of gp] (s) {$\sem{\inlinecode{g} - \inlinecode{e}}(F_T)$};

      \node[rectangle,above=0.0cm of g] (gcontent) {\begin{tabular}{r}$t_1 :: (1, 2)$ \\ $t_2 :: (2, 3)$\end{tabular}};
      \node[rectangle,above=0.0cm of e] (econtent) {\begin{tabular}{r}$t_3 :: (2, 3)$\end{tabular}};
      \node[rectangle,below=0.0cm of s] (scontent) {\begin{tabular}{r}$t_1 :: (1, 2)$ \\ $t_2 \otimes (\ominus~ t_3) :: (2, 3)$\end{tabular}};

      \draw[->] (g) -- (s.north west);
      \draw[->] (e) -- (s.north east);
    \end{tikzpicture}
    \caption{\ram~ tagged semantics}
    \label{fig:negation-example-symbolic}
  \end{subfigure}
  \hfill
  \begin{subfigure}{0.31\textwidth}
    \centering
    \begin{tikzpicture}
      \node[draw] (g) {$\sem{\inlinecode{g}}(F_\texttt{mmp})$};
      \node[right=0.5cm of g] (gp) {};
      \node[draw,right=0.5cm of gp] (e) {$\sem{\inlinecode{e}}(F_\texttt{mmp})$};
      \node[draw,below=0.4cm of gp] (s) {$\sem{\inlinecode{g} - \inlinecode{e}}(F_\texttt{mmp})$};

      \node[rectangle,above=0.0cm of g] (gcontent) {\begin{tabular}{r}$0.9 :: (1, 2)$ \\ $0.9 :: (2, 3)$\end{tabular}};
      \node[rectangle,above=0.0cm of e] (econtent) {\begin{tabular}{r}$0.2 :: (2, 3)$\end{tabular}};
      \node[rectangle,below=0.0cm of s] (scontent) {\begin{tabular}{r}$0.9 :: (1, 2)$ \\ $\text{min}(0.9, 1 - 0.2) = 0.8 :: (2, 3)$\end{tabular}};

      \draw[->] (g) -- (s.north west);
      \draw[->] (e) -- (s.north east);
    \end{tikzpicture}
    \caption{\ram~with \inlinecodesmall{max-min-prob}}
    \label{fig:negation-example-maxminprob}
  \end{subfigure}
  \vspace{-5px}
  \caption{
    An example rule adapted from \secref{sec:overview} is compiled to a \ram~ rule with difference.
    \inlinecodesmall{g} and \inlinecodesmall{e} are abbreviated relation names.
    The graphs illustrate the evaluation of expression $\inlinecodesmall{g} - \inlinecodesmall{e}$ under different semantics.
  }
  \label{fig:sclram-diff-example}
  \Description{}
  \vspace{-0.1in}
\end{figure}
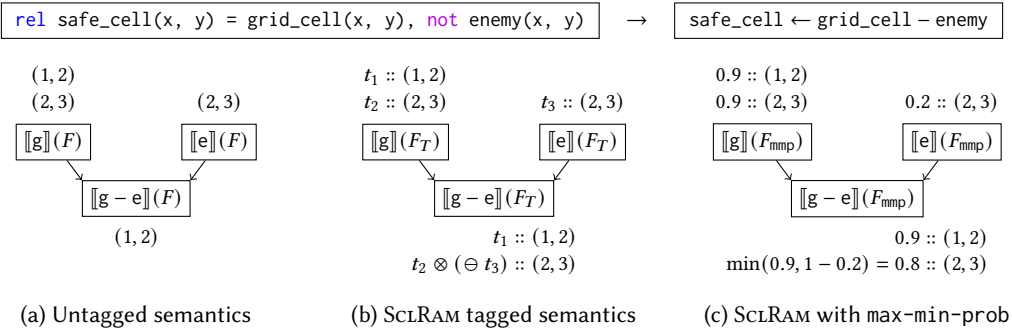

\textit{Difference and Negation.}
To evaluate a difference expression $e_1 - e_2$, there are two cases depending on whether a tuple $u$ evaluated from $e_1$ appears in the result of $e_2$.
If it does not, we simply propagate the tuple and its tag to the result ({\sc Diff-1});
otherwise, we get $t_1 :: u$ from $e_1$ and $t_2 :: u$ from $e_2$.
Instead of erasing the tuple $u$ from the result as in untagged semantics, we propagate a tag $t_1 \otimes (\ominus~ t_2)$ with $u$ ({\sc Diff-2}).
In this manner, information is not lost during negation.
\figref{fig:sclram-diff-example} compares the evaluations of an example difference expression under different semantics.
While the tuple $(2,3)$ is removed in the outcome with untagged semantics, it remains with the tagged semantics.

\begin{figure}
  \footnotesize

  \newcommand{\gridcell}[1]{\IfEqCase{#1}{{t}{\blacksquare}{f}{\Box}}}
  \newcommand{\gridbox}[9]{
    \scalebox{0.7}[0.7]{$
      \arraycolsep=0.0pt\def\arraystretch{0.3}
      \begin{array}{ccc}
        \gridcell{#1} & \gridcell{#2} & \gridcell{#3} \\
        \gridcell{#4} & \gridcell{#5} & \gridcell{#6} \\
        \gridcell{#7} & \gridcell{#8} & \gridcell{#9} \\
      \end{array}
    $}
  }

  \newcommand{\gridboxexample}{%
    \raisebox{1.5pt}{\scalebox{0.5}[0.5]{$
      \arraycolsep=0.0pt\def\arraystretch{0.3}
      \begin{array}{ccc}
        \blacksquare & \Box & \Box \\
        \Box & \Box & \Box \\
        \Box & \Box & \Box \\
      \end{array}
    $}}}

  \newcommand{\gridboxcorrect}{%
    \raisebox{1.5pt}{\scalebox{0.5}[0.5]{$
      \arraycolsep=0.0pt\def\arraystretch{0.3}
      \begin{array}{ccc}
        \Box & \Box & \Box \\
        \Box & \blacksquare & \blacksquare \\
        \Box & \Box & \Box \\
      \end{array}
    $}}}

  \begin{subfigure}{\textwidth}
    \centering
    \framebox[1.075\width]{\inlinescl{rel num\_enemies(n) = n := count(x, y: enemy(x, y))}}
    \quad
    $\rightarrow$
    \quad
    \raisebox{0.6pt}{\scalebox{0.90}{%
      \framebox[1.13\width]{$\inlinecode{num\_enemies} \leftarrow \pi_{\lambda n.(n)} ( \gamma_\texttt{count}(\inlinecode{enemy}) )$}%
    }}
  \end{subfigure}

  \vspace{5px}

  \begin{subfigure}{0.15\linewidth}
    \centering
    \includegraphics[width=0.9\linewidth]{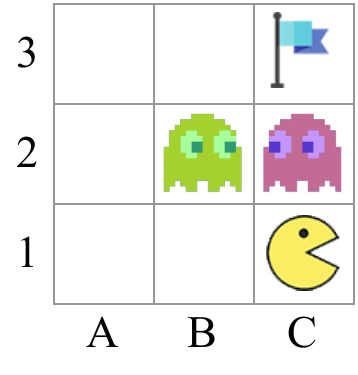}
    \caption{Maze image}
    \label{fig:sclram-running-example-1}
  \end{subfigure}
  \hfill
  \begin{subfigure}{0.15\linewidth}
    \centering
    \includegraphics[width=0.9\linewidth]{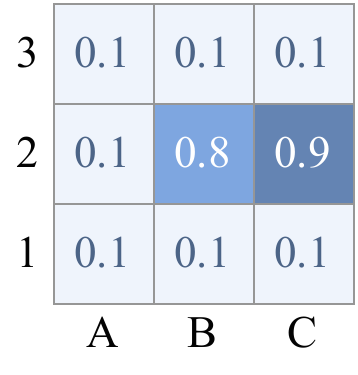}
    \caption{\inlinecodesmall{enemy}}
    \label{fig:sclram-running-example-2}
  \end{subfigure}
  \hfill
  \begin{subfigure}{0.67\linewidth}
    \centering
    \def\arraystretch{1.3}
    \begin{tabular}{cc}
      \scalebox{0.9}{\framebox{$\left\langle\sem{\gamma_\texttt{count}(\inlinecode{enemy})}(F_T)\right\rangle$}} &
      \scalebox{0.9}{\framebox{$\left\langle\sem{\gamma_\texttt{count}(\inlinecode{enemy})}(F_\texttt{mmp})\right\rangle$}} \\
      \gridbox{f}{f}{f}{f}{f}{f}{f}{f}{f} $:: 0$ & $0.1 :: 0$ \\
      \gridbox{t}{f}{f}{f}{f}{f}{f}{f}{f} $\oplus$ \gridbox{f}{t}{f}{f}{f}{f}{f}{f}{f} $\oplus \dots \oplus$ \gridbox{f}{f}{f}{f}{f}{f}{f}{t}{f} $\oplus$ \gridbox{f}{f}{f}{f}{f}{f}{f}{f}{t} $:: 1$ & $0.1 :: 1$ \\
      \gridbox{t}{t}{f}{f}{f}{f}{f}{f}{f} $\oplus$ \gridbox{t}{f}{t}{f}{f}{f}{f}{f}{f} $\oplus \dots \oplus$ {\color{sclblue}{\gridbox{f}{f}{f}{f}{t}{t}{f}{f}{f}}} $\oplus \dots \oplus$ \gridbox{f}{f}{f}{f}{f}{f}{t}{f}{t} $\oplus$ \gridbox{f}{f}{f}{f}{f}{f}{f}{t}{t} $:: 2$ & $0.8 :: 2$ \\
      \multicolumn{1}{c}{$\dots$} & $\dots$ \\
      \gridbox{t}{t}{t}{t}{t}{t}{t}{t}{t} $:: 9$ & $0.1 :: 9$
    \end{tabular}
    \caption{Evaluation of the aggregation expression}
    \label{fig:sclram-running-example-3}
  \end{subfigure}
  \vspace{-0.05in}
  \caption{
    An example counting enemies in a PacMan maze.
    Shown above are the \ours~ rule and compiled \ram~ rule with aggregation.
    (a) and (b) visualize the maze and the content of a probabilistic \inlinecodesmall{enemy} relation.
    For example, we have $t_{\texttt{B2}} :: \inlinecodesmall{enemy}(B, 2)$ where $t_{\texttt{B2}} = 0.8$.
    In (c), we show two normalized ($\left\langle . \right\rangle$ defined in \figref{fig:sclram-semantics}) evaluation results under abstract tagged semantics and with \inlinecodesmall{max-min-prob} provenance.
    Each symbol such as \gridboxexample~ represents a world corresponding to our arena ($\blacksquare$: enemy; $\Box$: no enemy).
    A world is a conjunction of 9 tags, e.g., $\gridboxexample  = t_{\texttt{A3}} \otimes (\ominus t_{\texttt{A2}}) \otimes \dots \otimes (\ominus t_{\texttt{C1}})$.
    We mark the correct world {\color{sclblue}{\gridboxcorrect}}~ which yields the answer $2$.
  }
  \label{fig:sclram-running-example}
  \Description{}
\end{figure}

\textit{Aggregation.}
Aggregators in \ram~ are discrete functions $g$ operating on set of (untagged) tuples $U \in \mathcal{U}$.
They return a {\it set} of aggregated tuples to account for aggregators like $\inlinecodesmall{argmax}$ which can produce multiple outcomes.
For example, we have $\inlinecodesmall{count}(U) = \{ |U| \}$.
However, in the probabilistic domain, discrete symbols do not suffice.
Given $n$ tagged tuples to aggregate over, each tagged tuple can be turned on or off, resulting in $2^n$ distinct \textit{worlds}.
Each world is a partition of the input set $U_T$ ($|U_T| = n$).
Denoting the positive part as $X_T$ and the negative part as $\overline{X}_T = U_T - X_T$, the tag associated with this world is a conjunction of tags in $X_T$ and negated tags in $\overline{X}_T$.
Aggregating on this world then involves applying aggregator $g$ on tuples in the positive part $X_T$.
This is inherently exponential if we enumerate all worlds.
However, we can optimize over each aggregator and each provenance to achieve better performance.
For instance, counting over \inlinecodesmall{max-min-prob} tagged tuples can be implemented by an $O(n \log(n))$ algorithm, much faster than exponential.
\figref{fig:sclram-running-example} demonstrates a running example and an evaluation of a counting expression under \inlinecodesmall{max-min-prob} provenance.
The resulting count can be 0-9, each derivable by multiple worlds.

\textit{Rules and Fixed-Point Iteration.}
Evaluating a rule $p \leftarrow e$ on database $F_T$ concerns evaluating the expression $e$ and merging the result with the existing facts under predicate $p$ in $F_T$.
The result of evaluating $e$ may contain duplicated tuples tagged by distinct tags, owing to expressions such as union, projection, or aggregation.
Therefore, we perform \textit{normalization} on the set to join ($\oplus$) the distinct tags.
From here, there are three cases to merge the newly derived tuples ($\left\langle\sem{e}(F_T)\right\rangle$) with the previously derived tuples ($\sem{p}(F_T)$).
If a fact is present only in the old or the new, we simply propagate the fact to the output.
When a tuple $u$ appears in both the old and the new, we propagate the disjunction of the old and new tag ($t^\old \oplus t^\new$).
Combining all cases, we obtain a set of newly tagged facts under predicate $p$.

\begin{figure}
  \footnotesize

  \newcommand{\cellsize}{0.12cm}

  \tikzset{
    dep q/.style={insert path={-- (-\cellsize, \cellsize)}},
    dep w/.style={insert path={-- (0, \cellsize)}},
    dep e/.style={insert path={-- (\cellsize, \cellsize)}},
    dep a/.style={insert path={-- (-\cellsize, 0)}},
    dep s/.style={insert path={-- (0, 0)}},
    dep d/.style={insert path={-- (\cellsize, 0)}},
    dep z/.style={insert path={-- (-\cellsize, -\cellsize)}},
    dep x/.style={insert path={-- (0, -\cellsize)}},
    dep c/.style={insert path={-- (\cellsize, -\cellsize)}},
    recurse lattice path/.code args={#1#2}{
      \ifx#1.\else\tikzset{dep #1,recurse lattice path=#2}\fi
    },
    lattice path/.style={recurse lattice path=#1.},
  }

  %
  %
\newcommand*{\gridpath}[1]{%
\raisebox{-4pt}{%
        \includegraphics[height=12pt,]{#1}%
}}

\newcommand*{\gridpathexample}{%
\raisebox{-3pt}{%
        \includegraphics[height=10pt,]{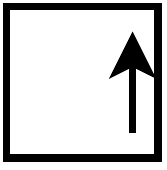}%
}}

\newcommand*{\gridpathgood}{%
\raisebox{-3pt}{%
        \includegraphics[height=10pt,]{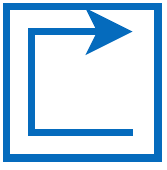}%
}}




  \begin{subfigure}{\textwidth}
    \centering
    \scalebox{0.88}{%
      \framebox[1.03\width]{\inlinescl{rel path(x1,y1,x3,y3) = (edge(x1,y1,x3,y3) or path(x1,y1,x2,y2) and edge(x2,y2,x3,y3)) and not enemy(x3,y3)}}%
    }
  \end{subfigure}

  \vspace{5px}

  \begin{subfigure}{\textwidth}
    \centering
    \def\arraystretch{1.4}
    \setlength{\tabcolsep}{0.6em}
    \begin{tabular}{r||c:c:c:c:c:c:c}
      Iteration count $i$ & 1 & 2 & 3 & 4 & 5 & 6 & 7 \\ \hline
      $t_{\texttt{C1-C3}}^{(i)}$ in $F_T^{(i)}$ & -- & \gridpath{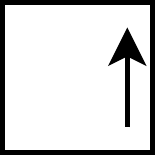} & (same) & $\gridpath{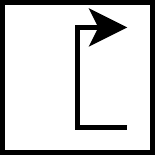} \oplus \gridpath{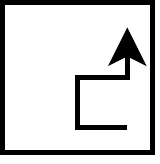} \oplus \cdots \oplus \gridpath{images/pacman/cde.pdf}$ & (same) & ${\gridpath{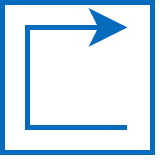}} \oplus \gridpath{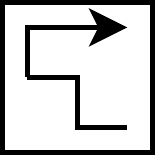} \oplus \cdots \oplus \gridpath{images/pacman/cde.pdf}$ & (same) \\
      $t_{\texttt{C1-C3}}^{(i)}$ in $F_{\texttt{mmp}}^{(i)}$ & -- & 0.1 & 0.1 & 0.2 & 0.2 & 0.9 & 0.9 \\ \hline
      $t_{\texttt{C1-C3}}^{(i)}$ saturated? & -- & false & true & false & true & false & true \\
      $F_{\texttt{mmp}}^{(i)}$ saturated? & false & false & false & false & false & false & true \\
    \end{tabular}
  \end{subfigure}
  \vspace{-8px}
  \caption{
    A demonstration of the fixed-point iteration to check whether actor at $\texttt{C1}$ can reach $\texttt{C3}$ without hitting an enemy (\figref{fig:sclram-running-example-1}).
    The \ours~ rule to derive this is defined on the top, and we assume bidirectional edges are populated and tagged by $\textbf{1}$.
    Let $t_{\texttt{C1-C3}}$ be the tag associated with $\inlinecodesmall{path(C,1,C,3)}$.
    We use a symbol like \gridpathexample~ to represent a conjunction of negated tags of \inlinecode{enemy} along the illustrated path, e.g. $\gridpathexample = (\ominus t_{\texttt{C2}}) \otimes (\ominus t_{\texttt{C3}})$.
    2nd iter is the first time $t_{\texttt{C1-C3}}$ is derived, but the path $\gridpathexample$ is blocked by an enemy.
    On 6th iter, the best path ${\color{sclblue}{\gridpathgood}}$ is derived in the tag.
    After that, under the \inlinecode{max-min-prob} provenance, both the tag $t_\texttt{C1-C3}$ and the database $F_\texttt{mmp}$ are saturated, causing the iteration to stop.
    Compared to untagged semantics in Datalog which will stop after 4 iterations, \ram~ with \inlinecode{mmp} saturates slower but allowing to explore better reasoning chains.
    %
  }
  \label{fig:sclram-recur-example}
  \Description{}
  \vspace{-0.15in}
\end{figure}

Recursion in \ram~ is realized similar to least fixed point iteration in Datalog \cite{abiteboul1995foundations}.
The iteration happens on a per-stratum basis to enforce stratified negation and aggregation.
Evaluating a single step of stratum $s$ means evaluating all the rules in $s$ and returning the updated database.
Note that we define a specialized least fixed point operator $\textbf{lfp}^\circ$, which stops the iteration once the whole database is \textit{saturated}.
\figref{fig:sclram-recur-example} illustrates an evaluation involving recursion and database saturation.
The whole database saturates on the 7th iteration, and finds the tag representing the optimal path for the PacMan to reach the goal.
Termination is not universally guaranteed in \ram~ due to the presence features such as value creation.
But its existence can be proven in a per-provenance basis.
For example, it is easy to show that if a program terminates under untagged semantics, then it terminates under tagged semantics with \inlinecodesmall{max-min-prob} provenance.

\subsection{External Interface and Execution Pipeline}
\label{sec:prov-extensions}

Thus far, we have only illustrated the \inlinecodesmall{max-min-prob} provenance, in which the tags are approximated probabilities.
There are other probabilistic provenances with more complex tags such as proof trees or boolean formulae.
We therefore introduce, for each provenance $T$, an \textit{input tag} space $I$, an \textit{output tag} space $O$, a \textit{tagging function} $\tau : I \rightarrow T$, and a \textit{recover function} $\rho : T \rightarrow O$.
For instance, all probabilistic provenances share the same input and output tag spaces $I = O = [0, 1]$ for a unified interface, while the internal tag spaces $T$ could be different.
We call the 4-tuple $(I, O, \tau, \rho)$ the \textit{external interface} for a provenance $T$.
The whole execution pipeline is then illustrated below:
$$
\begin{tikzpicture}[font=\footnotesize]
  \node[draw,minimum height=0.5cm] (edb) at (0,0) {$F_\texttt{option<$I$>}$};
  \node[draw,minimum height=0.5cm] (init-db) [right=1.2cm of edb] {$F_T$};
  \node[draw,minimum height=0.5cm] (result-db) [right=4.5cm of init-db] {$F_T'$};
  \node[draw,minimum height=0.5cm] (idb) [right=1.2cm of result-db] {$F_O$};
  \draw[->] (edb.east) -- (init-db.west) node [midway, fill=white] {$\tau$};
  \draw[->] (init-db.east) -- (result-db.west) node [midway, fill=white,draw,rounded corners=.23cm] {\hspace*{10pt}\ram~ program, $\overline{s}$\hspace*{10pt}};
  \draw[->] (result-db.east) -- (idb.west) node [midway, fill=white] {$\rho$};
\end{tikzpicture}
$$
In the context of a \ours~ application, an EDB is provided in the form $F_{\texttt{option<$I$>}}$.
During the \textit{tagging phase}, $\tau$ is applied to each input tag to obtain $F_T$, following which the \ram~ program operates on $F_T$.
For convenience, not all input facts need to be tagged---untagged input facts are simply associated by the tag $\textbf{1}$ in $F_T$.
In the \textit{recovery phase}, $\rho$ is applied to obtain $F_O$, the IDB that the whole pipeline returns.
\ours~ allows the user to specify a set of \textit{output relations} and $\rho$ is only applied to tags under such relations to avoid redundant computations.


\subsection{Differentiable Reasoning with Provenance}
\label{sec:prov-structures}

\begin{figure}
  \footnotesize

  \begin{minipage}{\textwidth}
    \centering
    \footnotesize
    \def\arraystretch{1.2}
    \setlength{\tabcolsep}{0.4em}
    \begin{tabular}{c||c|c|c|c|c|c|c||c|c}
      \textbf{Provenance}        & $T$ & $\textbf{0}$ & $\textbf{1}$ & $t_1 \oplus t_2$ & $t_1 \otimes t_2$ & $\ominus~ t$ & $t_1 \oeq t_2$ & $\tau(r_i)$ & $\rho(t)$ \\ \hline
      \texttt{diff-max-min-prob} & $\mathbb{D}$ & $\hat{0}$ & $\hat{1}$ & $\text{max}(t_1, t_2)$ & $\text{min}(t_1, t_2)$ & $\hat{1} - t$ & $t_1^{\text{fst}} ~\texttt{==}~ t_2^{\text{fst}}$ & $(r_i, \vec{\mathbf{e}}_i)$ & $t$ \\
      \texttt{diff-add-mult-prob} & $\mathbb{D}$ & $\hat{0}$ & $\hat{1}$ & $\text{clamp}(t_1 + t_2)$ & $t_1 \cdot t_2$ & $\hat{1} - t$ & true & $(r_i, \vec{\mathbf{e}}_i)$ & $t$ \\
      \texttt{diff-top-k-proofs} & $\Phi$ & $\emptyset$ & $\{\emptyset\}$ & $t_1 \vee_k t_2$ & $t_1 \wedge_k t_2$ & $\neg_k~ t$ & $t_1 ~\texttt{==}~ t_2$ & $\{\{\text{pos}(i)\}\}$ & $\text{WMC}(t, \Gamma)$ \\
    \end{tabular}
    \captionof{figure}{
      Definitions of three differentiable provenances.
    }
    \label{fig:provenance-example}
  \end{minipage}

  \begin{minipage}{\textwidth}
    \footnotesize
    \[
      \arraycolsep=2pt
      \begin{array}{rcl}
        \hat{a}_i &=& (a_i, \nabla a_i) \in \mathbb{D} \\
        \hat{0} &=& (0, \vec{0}) \\
        \hat{1} &=& (1, \vec{0})
      \end{array}
      \qquad
      \begin{array}{rcl}
        \hat{a}_1 + \hat{a}_2 &=& (a_1 + a_2, \nabla a_1 + \nabla a_2) \\
        \hat{a}_1 \cdot \hat{a}_2 &=& (a_1 \cdot a_2, a_2 \cdot \nabla a_1 + a_1 \cdot \nabla a_2) \\
        -\hat{a}_1 &=& (-a_1, -\nabla a_1)
      \end{array}
      \qquad
      \begin{array}{rcl}
        \text{min}(\hat{a}_1, \hat{a}_2) &=& \hat{a}_i, ~\text{where}~ i = \text{argmin}_i (a_i) \\
        \text{max}(\hat{a}_1, \hat{a}_2) &=& \hat{a}_i, ~\text{where}~ i = \text{argmax}_i (a_i) \\
        \text{clamp}(\hat{a}_1) &=& (\text{clamp}_0^1(a_1), \nabla a_1)
      \end{array}
    \]
    \vspace{-10px}
    \captionof{figure}{Operations on dual-number $\mathbb{D} \triangleq [0, 1] \times \mathbb{R}^n$, where $n$ is the number of input probabilities.}
    \label{fig:dual-number-def}
  \end{minipage}

  \begin{minipage}{\textwidth}
    \footnotesize
    \[
      \arraycolsep=3pt
      \begin{array}{rrcl}
        \text{(Variable)} & i & \in & 1 \dots n \\
        \text{(Literal)} & \nu & ::= & \text{pos}(i) \sep \text{neg}(i) \\
        \text{(Proof)} & \eta & ::= & \{ \nu_1, \nu_2, \dots \} \\
        \text{(Formula)} & \Phi \ni \varphi & ::= & \{ \eta_1, \eta_2, \dots \}
      \end{array}
      \quad
      \begin{array}{rcl}
        \varphi_1 \vee_k \varphi_2 & = & \text{top}_k(\varphi_1 \cup \varphi_2) \\
        \varphi_1 \wedge_k \varphi_2 & = & \text{top}_k(\{~ \eta \sep (\eta_1, \eta_2) \in \varphi_1 \times \varphi_2, \eta = \eta_1 \cup \eta_2, \eta ~\text{no conflict} ~\}) \\
        \neg_k~ \varphi & = & \text{top}_k(\text{cnf2dnf}(\{ \{\neg \nu \sep \nu \in \eta \} \sep \eta \in \varphi \})) \\
        \Gamma(i) & = & (r_i, \vec{\mathbf{e}}_i)
      \end{array}
    \]
    \vspace{-0.15in}
    \captionof{figure}{Definitions used for \inlinecode{diff-top-k-proofs} provenance.}
    \label{fig:top-k-proofs-def}
  \end{minipage}
  \Description{}
\end{figure}

We now elucidate how provenance also supports differentiable reasoning.
Let all the probabilities in the EDB form a vector $\vec{r} \in \mathbb{R}^n$, and the probabilities in the resulting IDB form a vector $\vec{y} \in \mathbb{R}^m$.
Differentiation concerns deriving output probabilities $\vec{y}$ as well as the derivative $\nabla \vec{y} = \frac{\partial \vec{y}}{\partial \vec{r}} \in \mathbb{R}^{m \times n}$.

In \ours, one can obtain these using a \textit{differentiable provenance} (DP).
DPs share the same external interface---let the input tag space $I = [0, 1]$ and output tag space $O$ be the space of \textit{dual-numbers} $\mathbb{D}$ (\figref{fig:dual-number-def}).
Now, each input tag $r_i \in [0, 1]$ is a probability,
and each output tag $\hat{y}_j = (y_j, \nabla y_j)$ encapsulates the output probability $y_j$ and its derivative w.r.t. inputs, $\nabla y_j$.
From here, we can obtain our expected output $\vec{y}$ and $\nabla \vec{y}$ by stacking together $y_j$-s and $\nabla y_j$-s respectively.

\ours~ provides 8 configurable built-in DPs with different empirical advantages in terms of runtime efficiency, reasoning granularity, and performance.
In this section, we elaborate upon 3 simple but versatile DPs, whose definitions are shown in \figref{fig:provenance-example}.
In the following discussion, we use $r_i$ to denote the $i$-th element of $\vec{r}$, where $i$ is called a \textit{variable} (ID).
Vector $\vec{\mathbf{e}}_i \in \mathbb{R}^n$ is the standard basis vector where all entries are 0 except the $i$-th entry.

\subsubsection{\inlinecodesmall{diff-max-min-prob} (\inlinecodesmall{dmmp})}
This provenance is the differentiable version of \inlinecodesmall{mmp}.
When obtaining $r_i$ from an input tag, we transform it into a dual-number by attaching $\vec{\mathbf{e}}_i$ as its derivative.
Note that throughout the execution, the derivative will always have at most one entry being non-zero and, specifically, $1$ or $-1$.
The saturation check is based on equality of the probability part only, so that the derivative does not affect termination.
All of its operations can be implemented by algorithms with time complexity $O(1)$, making it extremely runtime-efficient.

\subsubsection{\inlinecodesmall{diff-add-mult-prob} (\inlinecodesmall{damp})}
This provenance has the same internal tag space, tagging function, and recover function as \inlinecodesmall{dmmp}.
As suggested by its name, its disjunction and conjunction operations are just $+$ and $\cdot$ for dual-numbers.
When performing disjunction, we clamp the real part of the dual-number obtained from performing $+$, while keeping the derivative the same.
The saturation function for \inlinecodesmall{damp} is designed to always returns true to avoid non-termination.
But this decision makes it less suitable for complex recursive programs.
The time complexity of operations in \inlinecodesmall{damp} is $O(n)$, which is slower than \inlinecodesmall{dmmp} is but still very efficient in practice.

\subsubsection{\inlinecodesmall{diff-top-k-proofs} (\inlinecodesmall{dtkp})}
This provenance extends the \textit{top-$k$ proofs} semiring originally proposed in \cite{huang2021scallop} to additionally support negation and aggregation.
Shown in Fig. \ref{fig:provenance-example} and \ref{fig:top-k-proofs-def}, the tags of \inlinecodesmall{dtkp} are boolean formulas $\varphi \in \Phi$ in \textit{disjunctive normal form} (DNF).
Each conjunctive clause in the DNF is called a \textit{proof} $\eta$.
A formula can contain at most $k$ proofs, where $k$ is a tunable hyper-parameter.
Throughout execution, boolean formulas are propagated with $\vee_k$, $\wedge_k$, and $\neg_k$, which resemble \textit{or}, \textit{and}, and \textit{not} on DNF formulas.
At the end of these computations, $\text{top}_k$ is applied to keep only $k$ proofs with the highest \textit{proof probability}:
\begin{align}
  \Pr(\eta) = \textstyle \prod_{\nu \in \eta} \Pr(\nu), \quad \Pr(\text{pos}(i)) = r_i, \quad \Pr(\text{neg}(i)) = 1 - r_i.
\end{align}
When merging two proofs during $\wedge_k$, there might be conflicting literals, e.g. $\text{pos}(i)$ and $\text{neg}(i)$, in which case we remove the whole proof.
To take negation $\neg_k$ on $\varphi$, we first negate all the literals to obtain a \textit{conjunctive normal form} (CNF) equivalent to $\neg \varphi$.
Then we perform $\text{cnf2dnf}$ operation (conflict check included) to convert it back to a DNF.
To obtain the output dual-number $\hat{y}_j$ from a DNF formula $\varphi_j$, the tag for $j$-th output tuple, we adopt a differentiable \textit{weighted-model-counting} (WMC) procedure \cite{manhaeve2021deepproblog}.
WMC computes the weight of a boolean formula $\varphi_j$ given the weights of individual variables.
Concretely, $\hat{y}_j = \text{WMC}(\varphi_j, \Gamma)$ where $\Gamma(i) = (r_i, \vec{\mathbf{e}}_i)$ is the mapping from variables to their dual-numbers.
Note that WMC is \#P-complete, and is the main contributor to the runtime when using this provenance.
The tunable $k$ enables the user to balance between runtime and reasoning granularity.
Detailed runtime analysis is shown in Appendix \ref{app:sclram}.


\subsection{Practical Considerations}
\label{sec:prov-discussions}

We finally discuss some practical aspects concerning \ram~ extensions and provenance selection.

\textit{Additional Features.}
We only presented the syntax and semantics of the core fragment of \ram.
\ram~ additionally supports the following useful features:
1) sampling operations, and the provenance extension supporting them,
2) group-by aggregations,
3) tag-based early removal and its extension in provenance, and
4) mutually exclusive tags in \inlinecodesmall{dtkp}.
We formalize these in Appendix \ref{app:sclram}.


\textit{Provenance Selection.}
A natural question is how to select a differentiable provenance for a given \ours~ application.
Based on our empirical evaluation in \secref{sec:evaluation}, \inlinecodesmall{dtkp} is often the best performing one, and setting $k = 3$ is usually a good choice for both runtime efficiency and learning performance.
This suggests that a user can start with \inlinecodesmall{dtkp} as the default.
In general, the provenance selection process is similar to the process of hyperparameter tuning common in machine learning.

\section{Implementation}
\label{sec:implementation}

\begin{wrapfigure}{r}{0.2\textwidth}
  \footnotesize
  \centering
  \vspace{-10px}
  \begin{tabular}{c||c}
    \textbf{Module} & \textbf{LoC} \\ \hline
    Compiler & 19K \\ \hline
    Runtime & 16K \\ \hline
    Interpreter & 2K \\ \hline
    \ourpybind & 4K \\
  \end{tabular}
  \captionof{table}{LoC of core \ours~ modules.}
  \label{tab:scallop-module-loc}
  \vspace{-20px}
\end{wrapfigure}

We implemented the core \ours~ system in 45K LoC of Rust.
The LoC of individual modules is shown in \tabref{tab:scallop-module-loc}.
Within the compiler, there are two levels of intermediate representations, \textsf{front-IR} and \textsf{back-IR}, between the surface language and the \ram~ language.
In \textsf{front-IR}, we perform analyses such as type inference and transformations such as desugaring.
In \textsf{back-IR}, we generate query plans and apply optimizations.
The runtime operates directly on \ram~ and is based on semi-naive evaluation specialized for tagged semantics.
There are \textit{dynamic} and \textit{static} runtimes for interpreted and compiled \ram~ programs.
Currently, all computation is on CPU only, but can be parallelized per-batch for machine learning.

\ours~ can be used through different interfaces such as interpreter, compiler, and interactive terminal.
It also provides language bindings such as {\small\ourpybind}~ for Python and \ourwasmbind~ for WebAssembly and JavaScript.
With {\small\ourpybind}, \ours~ can be seamlessly integrated with machine learning frameworks such as PyTorch, wherein
the {\small\ourpybind}~ module is treated just like any other PyTorch module.
When \inlinecodesmall{jit} is specified in {\small\ourpybind}, \ours~ programs can be \textit{just-in-time} (JIT) compiled to Rust, turned into binaries, and dynamically loaded into the Python environment.

\ours~ provides 18 built-in provenances (4 for discrete reasoning, 6 for probabilistic, and 8 for differentiable).
The WMC algorithm is implemented using \textit{sentential decision diagram} (SDD) \cite{darwiche2011sdd} with naive bottom-up compilation.
We allow each provenance to provide its own implementation for operations such as aggregation and sampling.
This gives substantial freedom and enables optimizations for complex operations.
Our provenance framework is also interfaced with \ourpybind, allowing to quickly create and test new provenances in Python.

\section{Evaluation}
\label{sec:evaluation}

We evaluate the \ours~ language and framework on a benchmark suite comprising eight neurosymbolic applications.
Our evaluation aims to answer the following research questions:
\begin{itemize}
  \item[\textbf{RQ1}] How expressive is \ours~ for solving diverse neurosymbolic tasks?
  \item[\textbf{RQ2}] How do \ours's solutions compare to state-of-the-art baselines in terms of accuracy?
  \item[\textbf{RQ3}] Is the differentiable reasoning module of \ours~ runtime-efficient?
  \item[\textbf{RQ4}] Is \ours~ effective at improving generalizability, interpretability, and data-efficiency?
  \item[\textbf{RQ5}] What are the failure modes of \ours~ solutions and how can we mitigate them?
\end{itemize}
In the following sections, we first introduce the benchmark tasks and the chosen baselines for each task (\secref{sec:eval-bench}).
Then, we answer \textbf{RQ1} to \textbf{RQ5} in \secref{sec:eval-expressivity} to \secref{sec:eval-failure} respectively.
All \ours~ related and runtime related experiments were conducted on a machine with two 20-core Intel Xeon CPUs, four GeForce RTX 2080 Ti GPUs, and 768 GB RAM.

\subsection{Benchmarks and Baselines}
\label{sec:eval-bench}

\begin{figure}
  \centering
  \footnotesize
  \includegraphics[width=\textwidth]{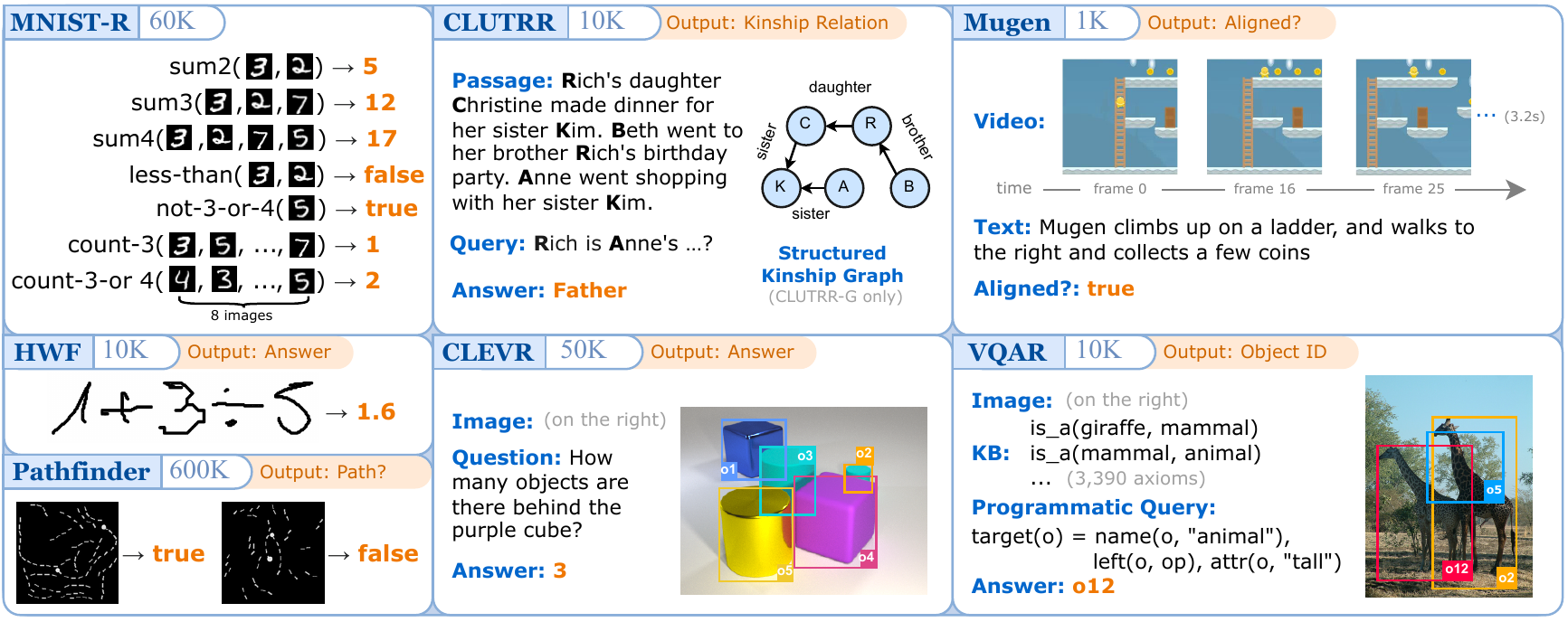}
 \vspace{-0.2in}
 \caption{
    Visualization of benchmark tasks.
    Beside the name of each task we specify the size of the training dataset and the output domain.
    PacMan-Maze is omitted since it has been shown in \secref{sec:overview}.
  }
  \label{fig:all-benchmarks}
  \Description{}
  \vspace{-0.1in}
\end{figure}

We present an overview of our benchmarks in \figref{fig:all-benchmarks}.
They cover a wide spectrum of tasks involving perception and reasoning.
The input data modality ranges from images and videos to natural language texts and knowledge bases (KB).
The size of the training dataset is also presented in the figure.
We next elaborate on the benchmark tasks and their corresponding baselines.

\textbf{\textit{MNIST-R}}: \textit{A Synthetic MNIST Test Suite.}
This benchmark is designed to test various features of \ours\ such as negation and aggregation.
Each task takes as input one or more images of hand-written digits from the MNIST dataset \cite{lecun1998mnist} and performs
simple arithmetic (sum2, sum3, sum4),
comparison (less-than),
negation (not-3-or-4),
or counting (count-3, count-3-or-4)
over the depicted digits.
For count-3 and count-3-or-4, we count digits from a set of 8 images.
For this test suite, we use a CNN-based model, DeepProbLog (DPL) \cite{manhaeve2021deepproblog}, and our prior work \cite{huang2021scallop} (Prior) as the baselines.


\textbf{\textit{HWF}}: \textit{Hand-Written Formula Parsing and Evaluation.}
HWF, proposed in \cite{li2020closed}, concerns parsing and evaluating hand-written formulas.
The formula is provided in the form of a sequence of images, where each image represents either a digit (0-9) or an arithmetic symbol ($+, -, \times, \div$).
Formulas are well-formed according to a grammar and do not divide by zero.
The size of the formulas ranges from 1 to 7 and is indicated as part of the input.
The goal is to evaluate the formula to obtain a rational number as the result.
We choose from \cite{li2020closed} the baselines NGS-$m$-BS, NGS-RL, and NGS-MAPO, which are \textit{neurosymbolic methods} designed specifically for this task.

\textbf{\textit{Pathfinder}}: \textit{Image Classification with Long-Range Dependency.}
In this task from \cite{tay2020longrange}, the input is an image containing two dots that are possibly connected by curved and dashed lines.
The goal is to tell whether the dots are connected.
There are two subtasks, Path and Path-X, where Path contains $32 \times 32$ images and Path-X contains $128 \times 128$ ones.
We pick as baselines standard CNN and Transformer based models, as well as the state-of-the-art neural models S4 \cite{gu2021S4}, S4$^*$ \cite{gu2022S4}, and SGConv \cite{li2022SGConv}.

\textbf{\textit{PacMan-Maze}}: \textit{Playing PacMan Maze Game.}
As presented in Section \ref{sec:overview}, this task tests an agent's ability to recognize entities in an image and plan the path for the PacMan to reach the goal.
An RL environment provides the game state image as input and the agent must plan the optimal action $\{\inlinecode{up}, \inlinecode{down}, \inlinecode{left}, \inlinecode{right}\}$ to take at each step.
There is no ``training dataset'' as the environment is randomized for every session.
We pick as baseline a CNN based Deep-Q-Network (DQN).
Unlike other tasks, we use the ``success rate'' metric for evaluation, i.e., among 1000 game sessions, we measure the number of times the PacMan reaches the goal within a certain time-budget.

\textbf{\textit{CLUTRR}}: \textit{Kinship Reasoning from Natural Language Context.}
In this task from \cite{sinha2019clutrr}, the input contains a natural language (NL) passage about a set of characters.
Each sentence in the passage hints at kinship relations.
The goal is to infer the relationship between a given pair of characters.
The target relation is not stated explicitly in the passage and it must be deduced through a reasoning chain.
Our baseline models include RoBERTa \cite{liu2019roberta}, BiLSTM \cite{graves2013bilstm}, GPT-3-FT (fine-tuned), GPT-3-ZS (zero-shot), and GPT-3-FS (5-shot) \cite{brown2020gpt3}.
In an alternative setup, CLUTRR-G, instead of the NL passage, the structured kinship graph corresponding to the NL passage is provided, making it a \textit{Knowledge Graph Reasoning} problem.
For CLUTRR-G, we pick GAT \cite{velivckovic2017gat} and CTP \cite{minervini2020ctp} as baselines.

\textbf{\textit{Mugen}}: \textit{Video-Text Alignment and Retrieval.}
Mugen \cite{hayes2022mugen} is based on a game called CoinRun \cite{cobbe2019coinrun}.
In the video-text alignment task, the input contains a 3.2 second long video of gameplay footage and a short NL paragraph describing events happening in the video.
The goal is to compute a similarity score representing how ``aligned'' they are.
There are two subsequent tasks, Video-to-Text Retrieval (VTR) and Text-to-Video Retrieval (TVR).
In TVR, the input is a piece of text and a set of 16 videos, and the goal is to retrieve the video that best aligns with the text.
In VTR, the goal is to retrieve text from video.
We compare our method with SDSC \cite{hayes2022mugen}.

\textbf{\textit{CLEVR}}: \textit{Compositional Language and Elementary Visual Reasoning \cite{clevr}.}
In this visual question answering (VQA) task, the input contains a rendered image of geometric objects and a NL question that asks about counts, attributes, and relationships of objects.
The goal is to answer the question based on the image.
We pick as baselines NS-VQA \cite{yi2018nsvqa} and NSCL \cite{mao2019nscl}, which are \textit{neurosymbolic methods} designed specifically for this task.

\textbf{\textit{VQAR}}: \textit{Visual-Question-Answering with Common-Sense Reasoning.}
This task, like CLEVR, also concerns VQA but with three salient differences:
it contains real-life images from the GQA dataset \cite{hudson2019gqa};
the queries are in a programmatic form, asking to retrieve objects in the image; and
there is an additional input in the form of a common-sense knowledge base (KB) \cite{gao2019cric} containing triplets such as (giraffe, is-a, animal) for common-sense reasoning.
The baselines for this task are NMNs \cite{andreas2016nmns} and LXMERT \cite{hao2019lxmert}.

\vspace{-5px}
\subsection{RQ1: Our Solutions and Expressivity}
\label{sec:eval-expressivity}

\begin{table}
  \footnotesize
  \setlength{\tabcolsep}{0.43em}
  \begin{tabular}{c||c|c|l|m{3.45cm}|c|c|c|c}
    \multirow{2}{*}{\textbf{Task}} & \multirow{2}{*}{\textbf{Input}} & \multirow{2}{*}{\textbf{Neural Net}} & \multirow{2}{*}{\ \ \ \ \ \textbf{Interface Relation(s)}} & \multirow{2}{*}{\ \ \ \ \ \ \ \ \ \ \ \textbf{\ours~ Program}} &
    \multicolumn{3}{c|}{\textbf{Features}} &
    \multirow{2}{*}{\textbf{LoC}} \\ \cline{6-8}
    & & & & & R & N & A & \\ \hline
    MNIST-R
    & Images & CNN
      & \inlinescl{digit(}\textit{id}, \textit{digit}\inlinescl{)}
        & Arithmetic, comparison, negation, and counting.
        & & \tick & \tick
        & 2$^\dagger$
      \\ \hline
    \multirow{2}{*}{HWF}
    & \multirow{2}{*}{Images} & \multirow{2}{*}{CNN}
      & \inlinescl{symbol(}\textit{id}, \textit{symbol}\inlinescl{)}
        & \multirow{2}{=}{Parses and evaluates formula over recognized symbols.}
        & \multirow{2}{*}{\tick} & &
        & \multirow{2}{*}{39}
      \\
    & & & \inlinescl{length(}\textit{len}\inlinescl{)} & & & & \\ \hline
    \multirow{2}{*}{Pathfinder}
    & \multirow{2}{*}{Image} & \multirow{2}{*}{CNN}
      & \inlinescl{dot(}\textit{id}\inlinescl{)}
        & \multirow{2}{=}{Checks if the dots are connected by dashes.}
        & \multirow{2}{*}{\tick} & &
        & \multirow{2}{*}{4}
      \\
    & & & \inlinescl{dash(}\textit{from\_id}, \textit{to\_id}\inlinescl{)} & & & & \\ \hline
    \multirow{3}{*}{\begin{tabular}{c}PacMan-\\Maze\end{tabular}}
    & \multirow{3}{*}{Image} & \multirow{3}{*}{CNN}
      & \inlinescl{actor(}\textit{x}, \textit{y}\inlinescl{)}
        & \multirow{3}{=}{Plans the optimal action by finding an enemy-free path from actor to goal.}
        & \multirow{3}{*}{\tick} & \multirow{3}{*}{\tick} & \multirow{3}{*}{\tick}
        & \multirow{3}{*}{31}
      \\
    & & & \inlinescl{enemy(}\textit{x}, \textit{y}\inlinescl{)} & & & & \\
    & & & \inlinescl{goal(}\textit{x}, \textit{y}\inlinescl{)} & & & & \\ \hline
    \multirow{3}{*}{\begin{tabular}{c}CLUTRR\\(-G)\end{tabular}}
    & NL & RoBERTa
      & \inlinescl{kinship(}\textit{rela}, \textit{sub}, \textit{obj}\inlinescl{)}
        & \multirow{3}{=}{Deduces queried relationship by recursively applying learnt composition rules.}
        & \multirow{3}{*}{\tick} & \multirow{3}{*}{\tick} & \multirow{3}{*}{\tick}
        & \multirow{3}{*}{8}
    \\ \cline{2-4}
    & Query$^*$ & -- & \inlinescl{question(}\textit{sub}, \textit{obj}\inlinescl{)} & & & & \\ \cline{2-4}
    & Rule & -- & \inlinescl{composition(}$r_1$, $r_2$, $r_3$\inlinescl{)} & & & & \\ \hline
    \multirow{3}{*}{Mugen}
    & Video & S3D
    & \inlinescl{action(}\textit{frame}, \textit{action}, \textit{mod}\inlinescl{)}
      & \multirow{3}{=}{Checks if events specified in NL text match the actions recognized from the video.}
      & \multirow{3}{*}{\tick} & \multirow{3}{*}{\tick} & \multirow{3}{*}{\tick}
      & \multirow{3}{*}{46}
    \\ \cline{2-4}
    & \multirow{2}{*}{NL} & \multirow{2}{*}{DistilBERT}
      & \inlinescl{expr(}\textit{expr\_id}, \textit{action}\inlinescl{)} & & & & \\
    & & & \inlinescl{mod(}\textit{expr\_id}, \textit{mod}\inlinescl{)} & & & & \\ \hline
    \multirow{4}{*}{CLEVR}
    & \multirow{2}{*}{Image} & \multirow{2}{*}{FastRCNN}
      & \inlinescl{obj\_attr(}\textit{obj\_id}, \textit{attr}, \textit{val}\inlinescl{)}
        & \multirow{4}{=}{Interprets CLEVR-DSL program (extracted from question) on scene graph (extracted from image).}
        & \multirow{4}{*}{\tick} & \multirow{4}{*}{\tick} & \multirow{4}{*}{\tick}
        & \multirow{4}{*}{51}
      \\
    & & & \inlinescl{obj\_rela(}\textit{rela}, $o_1$, $o_2$\inlinescl{)} & & & & \\ \cline{2-4}
    & \multirow{2}{*}{NL} & \multirow{2}{*}{BiLSTM}
      & \inlinescl{filter\_expr(}\textit{e}, \textit{ce}, \textit{attr}, \textit{val}\inlinescl{)} & & & & \\
    & & & \inlinescl{count\_expr(}\textit{e}, \textit{ce}\inlinescl{)}, \dots & & & & \\ \hline
    \multirow{4}{*}{VQAR}
    & \multirow{3}{*}{Image} & \multirow{3}{*}{FastRCNN}
      & \inlinescl{obj\_name(}\textit{obj\_id}, \textit{name}\inlinescl{)}
        & \multirow{4}{=}{Evaluates query over scene graphs (extracted from image) with the aid of common-sense knowledge base (KB).}
        & \multirow{4}{*}{\tick} & &
        & \multirow{4}{*}{42}
      \\
    & & & \inlinescl{obj\_attr(}\textit{obj\_id}, \textit{val}\inlinescl{)} & & & & \\
    & & & \inlinescl{obj\_rela(}\textit{rela}, $o_1$, $o_2$\inlinescl{)} & & & & \\ \cline{2-4}
    & KB$^*$
      & -- & \inlinescl{is\_a(}\textit{name1}, \textit{name2}\inlinescl{)}, \dots & & & & \\ \hline
  \end{tabular}
  \caption{
    Characteristics of \ours~ solutions for each task.
    Structured input which is not learnt is denoted by $^*$.
    Neural models used are RoBERTa \cite{liu2019roberta}, DistilBERT \cite{sanh2019distilbert}, and BiLSTM \cite{graves2013bilstm} for natural language (NL), CNN and FastRCNN \cite{girshick2015fastrcnn} for images, and S3D \cite{xie2018s3d} for video.
    We show the three key features of \ours~ used by each solution: (R)ecursion, (N)egation, and (A)ggregation.
    $\dagger$: For MNIST-R, the LoC is 2 for every subtask.
  }
  \vspace{-15px}
  \label{tab:benchmark-scallop-solution}
\end{table}

To answer \textbf{RQ1}, we demonstrate our \ours~ solutions to the benchmark tasks (\tabref{tab:benchmark-scallop-solution}).
For each task, we specify the interface relations which serve as the bridge between the neural and symbolic components.
The neural modules process the perceptual input and their outputs are mapped to (probabilistic) facts in the interface relations.
Our \ours~ programs subsequently take these facts as input and perform the described reasoning to produce the final output.
As shown by the \textit{features} column, our solutions use all of the core features provided by \ours.

The complete \ours~ program for each task is provided in the Appendix \ref{app:evaluation}.
These programs are succinct, as indicated by the LoCs in the last column of \tabref{tab:benchmark-scallop-solution}.
We highlight three tasks, HWF, Mugen, and CLEVR, to demonstrate \ours's expressivity.
For HWF, the \ours~ program consists of a formula parser.
It is capable of parsing probabilistic input symbols according to a context free grammar for simple arithmetic expressions.
For Mugen, the \ours~ program is a \textit{temporal specification checker}, where the specification is extracted from NL text to match the sequential events excerpted from the video.
For CLEVR, the \ours~ program is an interpreter for CLEVR-DSL, a domain-specific functional language introduced in the CLEVR dataset \cite{clevr}.

We note that our prior work \cite{huang2021scallop}, which only supports positive Datalog, cannot express 5 out of the 8 tasks since they need negation and aggregation, as indicated by columns `N' and `A'.
Moreover, HWF requires floating point support which is also lacking in our prior work.

Besides diverse kinds of perceptual data and reasoning patterns, the \ours~ programs are applied in a variety of learning settings.
As shown in \secref{sec:overview}, the program for PacMan-Maze is used in a \textit{online representation learning} setting.
For CLUTRR, we write integrity constraints (similar to the one shown in Section \ref{sec:language-horn-rules}) to derive \textit{semantic loss} used for constraining the language model outputs.
For CLUTRR-G, learnable weights are attached to \inlinecodesmall{composition} facts such as \inlinecodesmall{composition(FATHER, MOTHER, GRANDMOTHER)}, which enables to learn such facts
from data akin to \textit{rule learning} in ILP.
For Mugen, our program is trained in a \textit{contrastive learning} setup, since it requires to maximize similarity scores between aligned video-text pairs but minimize that for un-aligned ones.

\subsection{RQ2: Performance and Accuracy}
\label{sec:eval-accuracy}

\begin{figure}
    \begin{minipage}[c]{0.60\linewidth}
        \centering
        \footnotesize
        \begin{tikzpicture}
    \begin{axis}[
        ybar=0pt,
        width=1.02\textwidth,
        height = 4.2cm,
        ymin=0,
        ymax=106,
        bar width=1.1mm,
        ymajorgrids=true,
        major grid style={dotted,black},
        ylabel={Accuracy (\%)},
        ytick={0,25,50,...,100},
        y label style={at={(axis description cs:0.09,.5)}},
        symbolic x coords={sum2,sum3,sum4,less-than,not-3-or-4, count-3, count-3-or-4},
        xtick=data,
        x tick label style={yshift=1.5mm, rotate=-17},
        legend columns=3,
        legend style={
          draw=none,
          column sep=1ex,
        },
        legend style={at={(3.2cm,1)},anchor=south},
    ]

\addplot[color=mydeepblue,
        pattern color=mydeepblue,
        preaction={fill=myblue!50!white},
        pattern=north west lines]
coordinates{
    (sum2, 95.28)
    (sum3, 48.99)
    (sum4, 13.08)
    (less-than, 96.22)
    (not-3-or-4, 80.08)
    (count-3, 83.56)
    (count-3-or-4, 57.92)
};

\addplot[color=mydeepblue,
        pattern color=mydeepblue,
        preaction={fill=myblue!25!white},
        pattern=north east lines]
coordinates{
    (sum2, 96.82)
    (sum3, 94.90)
    (sum4, 0)
    (less-than, 98.04)
    (not-3-or-4, 68.26)
    (count-3, 0)
    (count-3-or-4, 0)
};

\addplot[color=mydeepblue,
        pattern color=mydeepblue,
        preaction={fill=myblue},
        pattern=horizontal lines]
coordinates{
    (sum2, 96.90)
    (sum3, 95.76)
    (sum4, 95.47)
    (less-than, 95.47)
    (not-3-or-4, 0)
    (count-3, 0)
    (count-3-or-4, 0)
};

\addplot[mydeeporange,fill=myorange!25!white]
coordinates{
    (sum2, 98.22)
    (sum3, 96.73)
    (sum4, 97.00)
    (less-than, 78.60)
    (not-3-or-4, 99.53)
    (count-3, 98.90)
    (count-3-or-4, 95.68)
};

\addplot[mydeeporange, fill=myorange!50!white]
coordinates{
    (sum2, 98.16)
    (sum3, 96.73)
    (sum4, 96.28)
    (less-than, 97.20)
    (not-3-or-4, 99.55)
    (count-3, 97.76)
    (count-3-or-4, 95.88)
};

\addplot[mydeeporange, fill=myorange!75!white]
coordinates{
    (sum2, 97.64)
    (sum3, 96.64)
    (sum4, 96.08)
    (less-than, 90.58)
    (not-3-or-4, 99.55)
    (count-3, 98.80)
    (count-3-or-4, 96.40)
};

\addplot[mydeeporange, fill=myorange]
coordinates{
    (sum2, 98.08)
    (sum3, 96.70)
    (sum4, 96.40)
    (less-than, 90.72)
    (not-3-or-4, 99.55)
    (count-3, 99.05)
    (count-3-or-4, 97.04)
};

\legend{
    CNN, 
    DPL, 
    Prior, 
    Ours (\inlinecode{dmmp}), 
    Ours (\inlinecode{damp}), 
    Ours (\inlinecode{dtkp-3}), 
    Ours (\inlinecode{dtkp-5}), 
}
\end{axis}
\end{tikzpicture}
        \vspace{-10px}
        \captionof{figure}{MNIST-R suite accuracy comparison.}
        \label{fig:mnist-overall}
        \Description{}
    \end{minipage}
    \hfill
    \begin{minipage}[c]{0.38\linewidth}
        \begin{minipage}[t]{\linewidth}
            \centering
            \footnotesize
            \setlength{\tabcolsep}{0.4em}
            \begin{tabular}{c||c|c|c|c}
                \multirow{2}{*}{\textbf{Method}} & \multicolumn{3}{c|}{\textbf{\ours}} & \multirow{2}{*}{\textbf{DQN}} \\ \cline{2-4}
                & \texttt{dmmp} & \texttt{damp} & \texttt{dtkp-1} & \\ \hline
                Succ Rate & 8.80\% & 7.84\% & \textbf{99.40\%} & 84.90\% \\ \hline
                \#Episodes & 50 & 50 & \textbf{50} & 50K
            \end{tabular}
            \captionof{table}{PacMan-Maze performance.}
            \label{tab:pacman-success-rate}
        \end{minipage}

        \vspace{-6px}

        \begin{minipage}[t]{\linewidth}
            \centering
            \footnotesize
            \begin{tikzpicture}
    \begin{axis}[
        width=1.1\linewidth, 
        height=1.37in,
        grid=major, 
        grid style={dashed,gray!30}, 
        xlabel=Epochs, 
        ylabel=Accuracy (\%),
        legend style={at={(1,0)},anchor=south east, font=\tiny}, 
        x label style={at={(axis description cs:0.5,0.15)},anchor=north},
        y label style={at={(axis description cs:0.15,.5)}},
        no markers,
        every axis plot/.append style={line width=1pt},
    ]
    \addlegendentry{\ours~ (\texttt{dmmp})};
    \addplot[mydeeporange] table[x=epoch, y=diffminmaxprob, col sep=comma, ] {data/hwf-learning-curve.csv};
    \addlegendentry{\ours~ (\texttt{damp})};
    \addplot[mydeeppurple, dashed] table[x=epoch, y=diffaddmultprob, col sep=comma, ] {data/hwf-learning-curve.csv};
    \addlegendentry{\ours~ (\texttt{dtkp-3})};
    \addplot[mydeepgreen] table[x=epoch, y=difftopbottomkclauses3, col sep=comma] {data/hwf-learning-curve.csv};
    \addlegendentry{\ours~ (\texttt{dtkp-5})};
    \addplot[mydeepblue] table[x=epoch, y=difftopbottomkclauses5, col sep=comma] {data/hwf-learning-curve.csv};

    \end{axis}
\end{tikzpicture}
            \vspace{-12px}
            \captionof{figure}{HWF learning curve.}
            \label{fig:hwf-eval-curve}
        \end{minipage}
    \end{minipage}

    \vspace{2px}

    \begin{minipage}[c]{\linewidth}
        \centering
        \footnotesize
\footnotesize
\centering

\makeatletter
 \pgfplotsset{
    calculate offset/.code={
        \pgfkeys{/pgf/fpu=true,/pgf/fpu/output format=fixed}
        \pgfmathsetmacro\testmacro{(\pgfplotspointmeta           *10^\pgfplots@data@scale@trafo@EXPONENT@y)*\pgfplots@y@veclength)}
        \pgfkeys{/pgf/fpu=false}
         },
    every node near coord/.style={
        /pgfplots/calculate offset,
         yshift=-\testmacro,
      },
    name node/.style={
    every node near coord/.append style={
                  name=#1-\coordindex
                  }},
    bar group size/.style 2 args={
        /pgf/bar shift={%
                -0.5*(#2*\pgfplotbarwidth + (#2-1)*\pgfkeysvalueof{/pgfplots/bar group skip})  +
                (.5+#1)*\pgfplotbarwidth + #1*\pgfkeysvalueof{/pgfplots/bar group skip}},%
    },
    bar group skip/.initial=0pt,
}

\makeatother

\begin{tikzpicture}
    \begin{axis}[
        ybar,
        ytick={0,25,50,...,100},
        width=\textwidth,
        height = 4.5cm,
        ymin=0,
        ymax=105,
        bar width=2mm,
        ymajorgrids=true,
        major grid style={dotted,black},
        ylabel={Accuracy (\%)},
        y label style={at={(axis description cs:0.05,.5)}},
        symbolic x coords={HWF, Path, Path-X, CLUTRR, CLUTRR-G, Mugen-TVR, Mugen-VTR, CLEVR, VQAR, },
        x tick label style={yshift=1.3mm, rotate=-15,},
        enlarge x limits=0.08,
        node near coords style={
            anchor=east,
            rotate=-90,
            font=\tiny,
        },
        table/x=n,
    ]

\addplot[mydeepblue, fill=myblue, nodes near coords=placeholder, bar group size={0}{3}, text opacity=0,]
coordinates{
   (HWF, 0)
   (Path, 0)
   (Path-X, 0)
   (CLUTRR-G, 0)
   (CLUTRR, 0)
   (CLEVR, 0)
   (VQAR, 0)
   (Mugen-TVR, 0)
   (Mugen-VTR, 0)
};

\addplot[mydeepblue, fill=myblue, nodes near coords=NGS-RL, text=mydeepblue!50!black,  bar group size={0}{4}]
coordinates{
  (HWF, 3.40)
};

\addplot[mydeepblue, fill=myblue, nodes near coords=NGS-MAPO, text=mydeepblue!50!black, bar group size={1}{4}]
coordinates{
  (HWF, 71.70)
};

\addplot[mydeepblue, fill=myblue, nodes near coords=NGS-$m$-BS, text=mydeepblue!50!black,  bar group size={2}{4}]
coordinates{
  (HWF, 98.50)
};

\addplot[mydeeporange, fill=myorange, nodes near coords=Ours (\texttt{dtkp-5}), text=mydeeporange!50!black,  bar group size={3}{4}]
coordinates{
  (HWF, 97.85)
};

\addplot[mydeepblue, fill=myblue, nodes near coords=Transformer, text=mydeepblue!50!black, bar group size={0}{6}]
coordinates{
  (Path, 71.40)
  (Path-X, 49.80)
};

\addplot[mydeepblue, fill=myblue, nodes near coords=CNN, text=mydeepblue!50!black, bar group size={1}{6}]
coordinates{
  (Path, 83.54)
  (Path-X, 85.32)
};

\addplot[mydeepblue, fill=myblue, nodes near coords=S4, text=mydeepblue!50!black, bar group size={2}{6}]
coordinates{
  (Path, 86.05)
  (Path-X, 88.10)
};

\addplot[mydeepblue, fill=myblue, nodes near coords=S4*, text=mydeepblue!50!black, bar group size={3}{6}]
coordinates{
  (Path, 94.20)
  (Path-X, 96.35)
};

\addplot[mydeepblue, fill=myblue, nodes near coords=SGConv, text=mydeepblue!50!black, bar group size={4}{6}]
coordinates{
  (Path, 95.46)
  (Path-X, 97.83)
};

\addplot[mydeeporange, fill=myorange, nodes near coords=Ours (\texttt{dtkp-3}), text=mydeeporange!50!black, bar group size={5}{6}]
coordinates{
  (Path, 87.42)
  (Path-X, 89.46)
};

\addplot[mydeepblue, fill=myblue, nodes near coords=BiLSTM, text=mydeepblue!50!black, bar group size={0}{6}]
coordinates{
  (CLUTRR, 34.9)
};

\addplot[mydeepblue, fill=myblue, nodes near coords=RoBERTa, text=mydeepblue!50!black, bar group size={1}{6}]
coordinates{
  (CLUTRR, 34.5)
};

\addplot[mydeepblue, fill=myblue, nodes near coords=GPT-3-ZS,text=mydeepblue!50!black, bar group size={2}{6}]
coordinates{
  (CLUTRR, 26.4)
};

\addplot[mydeepblue, fill=myblue, nodes near coords=GPT-3-FS, text=mydeepblue!50!black, bar group size={3}{6}]
coordinates{
  (CLUTRR, 33.1)
};

\addplot[mydeepblue, fill=myblue, nodes near coords=GPT-3-FT, text=mydeepblue!50!black, bar group size={4}{6}]
coordinates{
  (CLUTRR, 34.3)
};

\addplot[mydeeporange, fill=myorange, nodes near coords=Ours (\texttt{dtkp-3}), text=mydeeporange!50!black, bar group size={5}{6}]
coordinates{
  (CLUTRR, 61.34)
};

\addplot[mydeepblue, fill=myblue, nodes near coords=GAT, text=mydeepblue!50!black, bar group size={0}{3}]
coordinates{
  (CLUTRR-G, 39.05)
};

\addplot[mydeepblue, fill=myblue, nodes near coords=CTP, text=mydeepblue!50!black, bar group size={1}{3}]
coordinates{
  (CLUTRR-G, 95.57)
};

\addplot[mydeeporange, fill=myorange, nodes near coords=Ours (\texttt{dtkp-3}), text=mydeeporange!50!black, bar group size={2}{3}]
coordinates{
  (CLUTRR-G, 98.81)
};

\addplot[mydeepblue, fill=myblue, nodes near coords=NS-VQA, text=mydeepblue!50!black, bar group size={0}{3}]
coordinates{
  (CLEVR, 99.8)
};

\addplot[mydeepblue, fill=myblue, nodes near coords=NSCL, text=mydeepblue!50!black, bar group size={1}{3}]
coordinates{
  (CLEVR, 99.8)
};

\addplot[mydeeporange, fill=myorange, nodes near coords=Ours (\texttt{dmmp}), text=mydeeporange!50!black, bar group size={2}{3}]
coordinates{
  (CLEVR, 95.4)
};

\addplot[mydeepblue, fill=myblue, nodes near coords=LXMERT, text=mydeepblue!50!black, bar group size={0}{4}]
coordinates{
  (VQAR, 62.56)
};

\addplot[mydeepblue, fill=myblue, nodes near coords=NMNs, text=mydeepblue!50!black, bar group size={1}{4}]
coordinates{
  (VQAR, 71.80)
};

\addplot[mydeepblue, fill=myblue, nodes near coords=Prior, text=mydeepblue!50!black, bar group size={2}{4}]
coordinates{
  (VQAR, 84.22)
};

\addplot[mydeeporange, fill=myorange, nodes near coords=Ours (\texttt{dtkp-10}), text=mydeeporange!50!black, bar group size={3}{4}]
coordinates{
  (VQAR, 84.24)
};

\addplot[mydeepblue, fill=myblue, nodes near coords=SDSC, text=mydeepblue!50!black, bar group size={0}{2}]
coordinates{
  (Mugen-TVR, 86.80)
  (Mugen-VTR, 87.00)
};

\addplot[mydeeporange, fill=myorange, nodes near coords=Ours (\texttt{damp}), text=mydeeporange!50!black, bar group size={1}{2}]
coordinates{
  (Mugen-TVR, 88.10)
  (Mugen-VTR, 90.12)
};

\end{axis}
\end{tikzpicture}
        \vspace{-14px}
        \captionof{figure}{
            Overall benchmark accuracy comparison.
            The best-performing provenance structure for our solution is indicated for each task.
            Among the shown tasks, \inlinecode{dtkp} performs the best on 6 tasks, \inlinecode{damp} on 2, and \inlinecode{dmmp} on 1.
        }
        \label{fig:benchmark_overall}
    \end{minipage}
\end{figure}

To answer \textbf{RQ2}, we evaluate the performance and accuracy of our methods in terms of two aspects:
1) the best performance of our solutions compared to existing baselines, and
2) the performance of our solutions with different provenance structures (\inlinecodesmall{dmmp}, \inlinecodesmall{damp}, \inlinecodesmall{dtkp} with different $k$).

We start with comparing our solutions against selected baselines on all the benchmark tasks, as shown in \figref{fig:mnist-overall}, \tabref{tab:pacman-success-rate}, and \figref{fig:benchmark_overall}.
First, we highlight two applications, PacMan-Maze and CLUTRR, which benefit the most from our solution.
For PacMan-Maze, compared to DQN, we obtain a 1,000$\times$ speed-up in terms of training episodes, and a near perfect success rate of 99.4\%.
Note that our solution encodes environment dynamics (i.e. game rules) which are unavailable and hard to incorporate in the DQN model.
For CLUTRR, we obtain a 25\% improvement over baselines, which includes GPT-3-FT, the state-of-the-art large language model fine-tuned on the CLUTRR dataset.
Next, for tasks such as HWF and CLEVR, our solutions attain comparable performance, even compared to neurosymbolic baselines NGS-$m$-BS, NSCL, and NS-VQA specifically designed for each task.
On Path and Path-X, our solution obtains a 4\% accuracy gain over our underlying model CNN and even outperforms a carefully designed transformer based model S4.

The performance of the \ours~ solution for each task depends on the chosen provenance structure.
As can be seen from \tabref{tab:pacman-success-rate} and Figs. \ref{fig:mnist-overall}--\ref{fig:benchmark_overall}, although \inlinecodesmall{dtkp} is generally the best-performing one, each presented provenance is useful, e.g., \inlinecodesmall{dtkp} for PacMan-Maze and VQAR, \inlinecodesmall{damp} for less-than (MNIST-R) and Mugen, and \inlinecodesmall{dmmp} for HWF and CLEVR.
Note that under positive Datalog, \ours's \inlinecodesmall{dtkp} is identical to \cite{huang2021scallop}, allowing us to achieve similar performance.
In conclusion, allowing configurable provenance helps tailor our methods to different applications.

\subsection{RQ3: Runtime Efficiency}
\label{sec:eval-runtime-efficiency}

We evaluate the runtime efficiency of \ours~ solutions with different provenance structures and compare it against baseline neurosymbolic approaches.
As shown in \tabref{tab:runtime-efficiency-comparison}, \ours~ achieves substantial speed-up over DeepProbLog (DPL) on MNIST-R tasks.
DPL is a probabilistic programming system based on Prolog using exact probabilistic reasoning.
As an example, on sum4, DPL takes 40 days to finish only 4K training samples, showing that it is prohibitively slow to use in practice.
On the contrary, \ours~ solutions can finish a training epoch (15K samples) in minutes without sacrificing testing accuracy (according to \figref{fig:mnist-overall}).
For HWF, \ours~ achieves comparable runtime efficiency, even when compared against the hand-crafted and specialized NGS-$m$-BS method.

Comparing among provenance structures, we see significant runtime blowup when increasing $k$ for \inlinecodesmall{dtkp}.
This is expected as increasing $k$ results in larger boolean formula tags, making the WMC procedure exponentially slower.
In practice, we find $k = 3$ for \inlinecodesmall{dtkp} to be a good balance point between runtime efficiency and reasoning granularity.
In fact, \inlinecodesmall{dtkp} generalizes DPL, as one can set an extremely large $k \geq 2^n$ ($n$ is the total number of input facts) for exact probabilistic reasoning.

\subsection{RQ4: Generalizability, Interpretability, and Data-Efficiency}
\label{sec:eval-interpretability}

\begin{figure}
  \begin{minipage}[t][][b]{0.59\textwidth}
    \centering
    \footnotesize
    \begin{tabular}{c||r|r|r|r|r}
      \multirow{2}{*}{\textbf{Task}} & \multicolumn{4}{c|}{\textbf{\ours}} & \multirow{2}{*}{\textbf{Baseline}\ \ \ \ \ } \\ \cline{2-5}
      & \inlinecode{dmmp} & \inlinecode{damp} & \inlinecode{dtkp-3} & \inlinecode{dtkp-10} & \\ \hline
      sum2 & \textbf{34} & 88 & 72 & 185 & 21,430 (DPL) \\
      sum3 & \textbf{34} & 
      \textbf{119} & 71 & 1,430 & 30,898 (DPL) \\
      sum4 & \textbf{34} & 154 & 77 & 4,329 & timeout (DPL) \\
      less-than & 35 & \textbf{42} & 34 & 43 & 2,540 (DPL) \\
      not-3-or-4 & 37 & \textbf{33} & \textbf{33} & \textbf{34} & 3,218 (DPL) \\ \hline
      HWF & 89 & 107 & \textbf{120} & 8,435 & 79 (NGS-$m$-BS) \\ \hline
      CLEVR & \textbf{1,964} & 1,618 & 2,325 & timeout  & -- \\
    \end{tabular}
    \vspace{2px}
    \captionof{table}{
      Runtime efficiency comparison on selected benchmark tasks.
      Numbers shown are average training time (sec.) per epoch.
      Our variants attaining the best accuracy are indicated in bold.
    }
    \label{tab:runtime-efficiency-comparison}
  \end{minipage}
  \hfill
  \begin{minipage}[t][][b]{0.38\textwidth}
    \centering
    \footnotesize
    \begin{tikzpicture}
      \begin{axis}[
        width=1.05\linewidth, 
        height=1.50in,
        grid=major, 
        grid style={dashed,gray!30}, 
        xlabel={$k$, length of reasoning chain},
        ylabel=Accuracy (\%),
        ytick={25,50,75,100},
        legend style={at={(-0.2,1)},draw=none,anchor=south west, font=\tiny}, 
        x label style={at={(axis description cs:0.5,0.15)},anchor=north},
        y label style={at={(axis description cs:0.12,.5)}},
        no markers,
        every axis plot/.append style={line width=1pt},
        legend columns=3,
      ]

      \addlegendentry{Ours};
      \addplot[mydeeporange] table[x=k, y=ours, col sep=comma, ] {data/clutrr-sys-generalizability.csv};
      \addlegendentry{BiLSTM};
      \addplot[mydeeppurple, dashed] table[x=k, y=bilstm, col sep=comma, ] {data/clutrr-sys-generalizability.csv};
      \addlegendentry{RoBERTa};
      \addplot[mydeepgreen, dashed] table[x=k, y=roberta, col sep=comma, ] {data/clutrr-sys-generalizability.csv};
      \addlegendentry{GPT-3-FT};
      \addplot[mydeepred, densely dotted] table[x=k, y=gpt3ft, col sep=comma] {data/clutrr-sys-generalizability.csv};
      \addlegendentry{GPT-3-FS};
      \addplot[mydeepblue, densely dotted] table[x=k, y=gpt3fs, col sep=comma] {data/clutrr-sys-generalizability.csv};

      \end{axis}
    \end{tikzpicture}

    \vspace{-10px}

    \captionof{figure}{
      Systematic generalizability on CLUTRR dataset.
    }
    \label{fig:clutrr-sys-generalizability}
  \end{minipage}
  
  \vspace{-1px}

  \begin{minipage}[t]{0.48\textwidth}
    \centering
    \footnotesize
    \begin{tabular}{ccc}
      \includegraphics[width=0.7in]{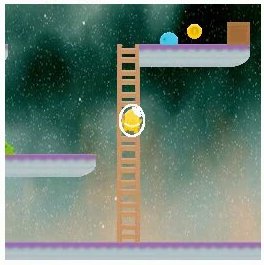} &
      \includegraphics[width=0.7in]{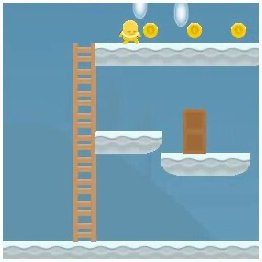} &
      \includegraphics[width=0.7in]{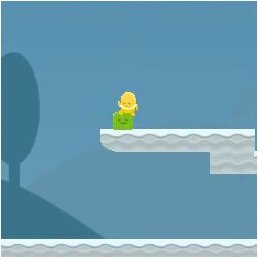} \\
      \inlinescl{(climb,up)} &
      \inlinescl{(collect,coin)} &
      \inlinescl{(kill,face)}
    \end{tabular}
    \vspace{-5px}
    \captionof{figure}{
      The predicted most likely (action, mod) pair for example video segments from Mugen dataset.
    }
    \label{fig:mugen-interpretability}
  \end{minipage}
  \hfill
  \begin{minipage}[t]{0.48\textwidth}
    \centering
    \footnotesize
    \begin{tabular}{c||c|c|c|c}
      \multirow{2}{*}{\textbf{\%Train}} & \textbf{\ours} & \multicolumn{3}{c}{\textbf{NGS}} \\ \cline{2-5}
      & \texttt{dtkp-5} & RL & MAPO & $m$-BS \\ \hline
      100\% & 97.85 & 3.4 & 71.7 & 98.5 \\ \hline
      50\% & 95.7 & 3.6 & 9.5 & 95.7 \\ \hline
      25\% & 92.95 & 3.5 & 5.1 & 93.3 \\
    \end{tabular}
    \vspace{1.5px}
    \captionof{table}{
      Testing accuracy of various methods on HWF when trained with only a portion of the data.
      Numbers are in percentage (\%).}
    \label{tab:hwf-data-efficiency}
  \end{minipage}
  \Description{}
  \vspace{-10px}
\end{figure}

We now consider other important desirable aspects of machine learning models besides accuracy and runtime, such as generalizability on unseen inputs, interpretability of the outputs, and data-efficiency of the training process.  For brevity, we focus on a single benchmark task in each case.

We evaluate \ours's generalization ability for the CLUTRR task.
Each data-point in CLUTRR is annotated with a parameter $k$ denoting the length of the reasoning chain to infer the target kinship relation.
To test different solutions' \textit{systematic generalizability}, we train them on data-points with $k \in \{2, 3\}$ and test on data-points with $k \in \{2, \dots, 10\}$.
As shown in \figref{fig:clutrr-sys-generalizability}, the neural baselines suffer a steep drop in accuracy on the more complex unseen instances, whereas the accuracy of \ours's solution degrades more slowly, indicating that it is able to generalize better.

Next, we demonstrate \ours's interpretability on the Mugen task.
Although the goal of the task is to see whether a video-text pair is aligned, the perceptual model in our method extracts interpretable symbols, i.e., the action of the controlled character at a certain frame.
\figref{fig:mugen-interpretability} shows that the predicted (action, mod) pairs perfectly match the events in the video.
Thus, our solution not only tells whether a video-text pair is aligned, but also {\em why} it is aligned.

Lastly, we evaluate \ours's data-efficiency on the HWF task, using lesser training data (\tabref{tab:hwf-data-efficiency}).
While methods such as NGS-MAPO suffer significantly when trained on less data, \ours's testing accuracy decreases slowly, and is comparable to the data-efficiency of the state-of-the-art neurosymbolic NGS-$m$-BS method.
PacMan-Maze task also demonstrates \ours's data-efficiency, as it takes much less training episodes than DQN does, while achieving much higher success rate.

\subsection{RQ5: Analysis of Failure Modes}
\label{sec:eval-failure}

Compared to purely neural models,
\ours~ solutions provide more transparency, allowing programmers to debug effectively.
By manually checking the interface relations, we observed that the main source of error lies in inaccurate predictions from the neural components.
For example, the RoBERTa model for CLUTRR correctly extracts only 84.69\% of kinship relations.
There are two potential causes---either the neural component is not powerful enough, or our solution is not providing adequate supervision to train it.
The former can be addressed by employing better neural architectures or more data.
The latter can be mitigated in different ways, such as tuning the selected provenance or incorporating integrity constraints (discussed in \secref{sec:language-horn-rules}) into training and/or inference.
For instance, in PacMan-Maze, a constraint that ``there should be no more than one \textit{goal} in the arena'' helps to improve robustness.




\section{Related Work}
\label{sec:related}

We survey related work in four different but overlapping domains:
provenance reasoning, Datalog and logic programming, probabilistic and differentiable programming, and neurosymbolic methods.

\textit{Relational Algebra and Provenance.}
Relational algebra is extensively studied in databases \cite{abiteboul1995foundations}, and extended with recursion \cite{jachiet20recursivequeries} in Datalog.
The provenance semiring framework \cite{provenancesemiring} was first proposed for positive relational algebra (RA$^{+}$) and later extended with difference \cite{provenanceondifference} and fixed-point \cite{semiring-provenance-for-fixed-point-logic}.
It is deployed in variants of Datalog \cite{mahmoud2022convergence-datalog} and supports applications such as program synthesis \cite{si2019difflog}.
\ours~employs an extended provenance semiring framework and demonstrates its application to configurable differentiable reasoning.

\textit{Datalog}
is a declarative logic programming language where rules are logical formulas.
Despite being less expressive than Prolog, it is widely studied in both the programming language and database communities for language extensions and optimizations \cite{mahmoud2022convergence-datalog}.
A variety of Datalog-based systems has been built for program analysis \cite{bernhard2016souffle} and enterprise databases \cite{molham2015logicblox}.
\ours~ is also a Datalog-based system that extends \cite{huang2021scallop} with numerous language constructs such as negation and aggregation.

\textit{Probabilistic Programming}
is a paradigm for programmers to model distributions and perform probabilistic sampling and inference.
Such systems include ProbLog \cite{anton2015problog2}, Pyro \cite{bingham2018pyro}, Turing \cite{ge2018t}, and PPL \cite{meent2018intro-ppl}.
When integrated with modern ML systems, they are well suited for statistical modeling and building generative models.
\ours~ also supports ProbLog-style exact probabilistic inference as one instantiation of our provenance framework.
But advanced statistical sampling and generative modeling is not yet supported and left for future work.

\textit{Differentiable Programming}
(DP) systems seek to enable writing code that is differentiable.
Common practices for DP include symbolic differentiation and automatic differentiation (auto-diff) \cite{bayd15autodiff}, resulting in popular ML frameworks such as PyTorch~\citep{paszke2019pytorch}, TensorFlow~\citep{abadi2015tensorflow}, and JAX~\citep{bradbury2018jax}.
However, most of these systems are designed for real-valued functions.
\ours~ is also a differentiable programming system, but with a focus on programming with discrete, logical, and relational symbols.

\textit{Neurosymbolic Methods}
have emerged to incorporate symbolic reasoning into existing learning systems.
Their success has been demonstrated in a number of applications \cite{yi2018nsvqa, mao2019nscl, li2020closed, wang2019satnet, xu2022dont, chen2020nerd, minervini2020ctp}.
Similar to \cite{yi2018nsvqa,mao2019nscl,li2020closed}, we focus primarily on solutions involving perception followed by symbolic reasoning.
Other ways of incorporating symbolic knowledge include \textit{semantic loss} \cite{xu18semanticloss,xu2022dont}, \textit{program synthesis} \cite{shah2020learning, chen2021webqa}, and invoking large language models (LLMs) \cite{cheng2022binding, zelikman2023parsel}.
Most aforementioned neurosymbolic methods build their own domain-specific language or specialized reasoning components, many of which are programmable in \ours~ and are instantiations of our provenance framework \cite{mao2019nscl,chen2020nerd,xu18semanticloss,xu2022dont}.
TensorLog \cite{cohen2017tensorlog}, DPL \cite{manhaeve2021deepproblog}, and \cite{huang2021scallop} can be viewed as unified neurosymbolic frameworks of varying expressivity.
\ours~ is inspired by DPL, and additionally offers a more scalable, customizable, and easy-to-use language and framework.

\section{Conclusion}
\label{sec:conclusion}

We presented \ours, a neurosymbolic programming language for integrating deep learning and logical reasoning.
We introduced a declarative language and a reasoning framework based on provenance computations.
We showed that our framework is practical by applying it to a variety of machine learning tasks.
In particular, our experiments show that \ours~ solutions are comparable and even supersede many existing baselines.

In the future, we plan to extend \ours~in three aspects:
1) Supporting more machine learning paradigms, such as generative modeling, open domain reasoning, in-context learning, and adversarial learning.
2)~Further enhancing the usability, efficiency, and expressiveness of \ours's language and framework.
We intend to provide bindings to other ML frameworks such as TensorFlow and JAX,
leverage hardware such as GPUs to accelerate computation,
and support constructs such as algebraic data types.
3) Applying \ours~to real-world and safety-critical domains.
For instance, we intend to integrate it with the CARLA driving simulator \cite{alexey2017carla} to specify soft temporal constraints for autonomous driving systems.
We also intend to apply \ours~ in the medical domain for explainable disease diagnosis from electronic health records (EHR) data.

\section*{Acknowledgement}

We thank Neelay Velingker, Hanjun Dai, Hanlin Zhang, and Sernam Lin for helpful comments on the presentation and experiments.
We thank the anonymous reviewers and artifact reviewers for useful feedback.
This research was supported by grants from DARPA (\#FA8750-19-2-0201), NSF (\#2107429 and \#1836936), and ONR (\#N00014-18-1-2021).

\section*{Artifact Availability}

All software necessary to reproduce the experiments in this paper is available at \cite{scallop-artifact}.
Additionally, the latest source code of Scallop and its documentation is available at \url{https://github.com/scallop-lang/scallop}.

\bibliography{index}

\appendix

\newpage
\section{\ours~ Language Syntax}
\label{app:scallop-surface-language}

\begin{figure}
  \centering
  \footnotesize
  \[
  \begin{array}{rrl}
  item &  :=  & \texttt{@} attrName ( attrArg * ) def \\
  def & := & importDef \sep typeDef \sep constDef \sep relaDef \sep queryDef \\
  importDef &  :=  & \code{import}~file \\
  typeDef &  :=  & \code{type}~typeName~\code{=}~aliasName \sep \code{type}~typeName~\code{<:}~superTypeName \\
           & \sep & \code{type}~relationName\code{(}[name \code{:}]type^*\code{)} \\
  constDef &  :=  & \code{const}~constName[\code{:}~type]~\code{=}~constant \\
  relaDef &  :=  & \code{rel}~[tag~\code{::}]~relaName\code{(}expr^*\code{)} \sep \code{rel}~relaName~\code{=}~\code{\{}taggedTuple^*\code{\}} \\
            & \sep & \code{rel}~[tag~\code{::}]~atom~\code{:-}~formula \\
  taggedTuple & := & [tag~\code{::}]~\code{(}constant^*\code{)} \\
  atom &  :=  & relaName\code{(}expr^*\code{)} \\
  formula &  :=  & formula~\code{and}~formula \sep formula~\code{or}~formula \sep formula~\code{implies}~formula \\
          & \sep & atom \sep \neg atom \sep constraint \sep reduce \\
  reduce &  :=  & var^*~\code{=}~aggregator(var^*\code{:}~formula~[\code{where}~var^*\code{:}~formula]) \\
          & \sep & var^*~\code{=}~sampler(var^*\code{:}~formula~[\code{where}~var^*\code{:}~formula]) \\
  constraint & := & expr \\
  type &  :=  & \code{u8} \sep \code{u16} \sep \code{u32} \sep \code{u64} \sep \code{usize} \sep \code{i8} \sep \code{i16} \sep \code{i32} \sep \code{i64} \sep \code{isize} \sep \code{f32} \sep \code{f64} \sep \code{bool} \sep \code{char} \sep \code{String} \\
  expr &  :=  & expr ~binOp~ expr \sep unaryOp~ expr \sep expr ~\code{as}~ type \sep \code{if}~expr~\code{then}~expr~\code{else}~expr \\
      & \sep & \$ func\code{(}expr^*\code{)} \sep variable \sep constant \\
  binOp &  :=  & \code{+} \sep \code{-} \sep \code{*} \sep \code{/} \sep \code{\%} \sep \code{\&\&} \sep \code{||} \sep \code{==} \sep \code{!=} \sep \code{<} \sep \code{<=} \sep \code{>} \sep \code{>=} \\
  unaryOp &  :=  & \code{!} \sep \code{-} \\
  aggregator &  :=  & \code{min} \sep \code{max} \sep \code{argmin<}var^*\code{>} \sep \code{argmax<}var^*\code{>} \sep \code{count} \sep \code{sum} \sep \code{prod} \sep \code{exists} \sep \code{forall} \\
  sampler & := & \code{top<K>} \sep \code{categorical<K>} \sep \code{uniform<K>}
  \end{array}
  \]
  \caption{Language grammar ($var: variable$ and $f: formula$).}
  \label{fig:language-grammar}
  \Description{}
\end{figure}

In this appendix, we provide the abstract syntax of the full language of \ours~ in \figref{fig:language-grammar}.

\section{Reasoning Framework}
\label{app:sclram}

In this appendix, we provide the extensions to our provenance framework, and illustrate the additional features provided by full \ram.

\subsection{Provenance Extensions}

\begin{figure}
  \footnotesize
  \[
    \begin{array}{rrcl}
    \text{(Tag)} & t & \in & T \\
    \text{(False)} & \mathbf{0} & \in & T \\
    \text{(True)} & \mathbf{1} & \in & T \\
    \text{(Disjunction)} & \oplus & : & T \times T \rightarrow T \\
    \text{(Conjunction)} & \otimes & : & T \times T \rightarrow T \\
    \text{(Negation)} & \ominus & : & T \rightarrow T \\
    \text{(Saturation)} & \oeq & : & T \times T \rightarrow \text{Bool} \\
    \text{(Early Removal)} & \text{discard} & : & T \rightarrow \text{Bool} \\
    \text{(Weight)} & \text{weight} & : & T \rightarrow \mathbb{R} \\
    \end{array}
  \]
  \vspace{-0.15in}
  \captionof{figure}{Full interface for a provenance structure.}
  \label{fig:full-provenance-interface}
  \Description{}
\end{figure}

We present the full provenance interface including two additional features, early removal and weight for sampling, in \figref{fig:full-provenance-interface}.
For early removal, we introduce a function \textit{discard} where it returns a boolean value based on a tag.
When returned true, the fact associated with this tag will be early removed from the computation.
For sampling, we allow each provenance to implement a function \textit{weight} that returns a real number in $\mathbb{R}$ representing the weight of the tag.
These features will be used later when we introduce the full \ram~ syntax and semantics.

\subsection{Abstract Syntax of \ram}

\begin{figure}
  \footnotesize
  \[
    \begin{array}{rcrl}
    \text{(Predicate)} & p \\
    \text{(Aggregator)} & g & ::= & \inlinecode{count} \sep \inlinecode{sum} \sep \inlinecode{prod} \sep \inlinecode{exists} \sep \inlinecode{max} \sep \inlinecode{min} \sep \inlinecode{argmax} \sep \inlinecode{argmin} \\
    \text{(Sampler)} & \mu & ::= & \inlinecode{top<K>} \sep \inlinecode{categorical<K>} \sep \inlinecode{uniform<K>} \\
    \text{(Expression)} & e & ::= & \varnothing \sep p \sep \mathbbm{1}(e) \sep \emptyset(e) \sep \gamma_g(e) \sep \hat{\gamma}_g(e_1, e_2) \sep \psi_\mu(e) \sep \hat{\psi}_\mu(e_1, e_2) \sep \pi_\alpha(e) \sep \sigma_\beta(e) \\
                        &   & \sep & e_1 \cup e_2 \sep e_1 \times e_2 \sep e_1 \bowtie e_2 \sep e_1 \cap e_2 \sep  e_1 - e_2 \\
    \text{(Rule)} & r & ::= & p \leftarrow e \\
    \text{(Stratum)} & s & ::= & \{ r_1, \dots, r_n \} \\
    \text{(Program)} & \overline{s} & ::= & s_1; \dots; s_n \\
    \end{array}
  \]
  \vspace{-0.15in}
  \captionof{figure}{Full abstract syntax of \ram.}
  \label{fig:full-sclram-syntax}
  \Description{}
\end{figure}

We revisit the abstract syntax of \ram, and add extensions including:
\begin{enumerate}
  \item Empty-set ($\varnothing$), One-overwrite ($\mathbbm{1}(e)$) and zero-overwrite ($\emptyset(e)$),
  \item Intersection ($\cap$), Natural-Join ($\bowtie$), and Anti-join ($\triangleright$),
  \item Group-by aggregate ($\hat{\gamma}_g(e_1, e_2)$), where $e_1$ is the expression finding groups, and $e_2$ is the main expression to aggregate over, conditioned on the groups found by evaluating $e_1$,
  \item Sampling ($\psi_\mu(e)$) where $\mu$ denotes a sampler,
  \item Group-by sampling ($\hat{\psi}_\mu(e_1, e_2)$), which is sampling but similar to aggregate, we have $e_1$ being the expression finding groups.
\end{enumerate}

\paragraph{Remark}

The One-overwrite $\mathbbm{1}(e)$ is particularly useful for defining magic-set transformation under tagged semantics.
Overwriting $\textbf{1}$ will make the tuples in magic-set predicate serve as pure demand fact for optimization, not conditioned on some given tags.

\subsection{Operational Semantics of \ram}

We show the full semantics in \figref{fig:full-sclram-semantics} and \figref{fig:full-sclram-semantics-2}, with the addition of group-by, sampling, and early-removal.
Note that the sampler $\mu : \mathcal{U}_\mathbb{R} \rightarrow \mathcal{U}$ takes in facts tagged by weights, and samples according to the weight.
The early removal happens in the stage of normalization before each rule update.

\begin{figure}
  \footnotesize
  \textbf{Expression semantics}
  \hfill
  \framebox[1.1\width]{$
    \alpha : \mathbb{U} \rightharpoonup \mathbb{U}, \quad
    \beta : \mathbb{U} \rightarrow \text{Bool},\quad
    g : \mathcal{U} \rightarrow \mathcal{U}, \quad
    \mu : \mathcal{U}_\mathbb{R} \rightarrow \mathcal{U}, \quad
    \sem{e} : \mathcal{F}_T \rightarrow \mathcal{U}_T
  $}
  $$
  \inferrule*[right=\anno{Empty Set}]
    { }
    {\sem{\varnothing}(F_T) = \emptyset}
  \quad
  \inferrule*[right=\anno{Predicate}]
    {t :: p(u) \in F_T}
    {t :: u \in \sem{p}(F_T)}
  $$
  $$
  \inferrule*[right=\anno{\textbf{0}-Overwrite}]
    {t :: u \in \sem{e}(F_T)}
    {\bm{0} :: u \in \sem{\emptyset(e)}(F_T)}
  \quad
  \inferrule*[right=\anno{\textbf{1}-Overwrite}]
    {t :: u \in \sem{e}(F_T)}
    {\bm{1} :: u \in \sem{\mathbbm{1}(e)}(F_T)}
  $$
  $$
  \inferrule*[right=\anno{Select}]
    {t :: u \in \sem{e}(F_T) \\ \beta(u) = \text{true}}
    {t :: u \in \sem{\sigma_\beta(e)}(F_T)}
  \hspace*{5pt}
  \inferrule*[right=\anno{Project}]
    {t :: u \in \sem{e}(F_T) \\ u' = \alpha(u)}
    {t :: u' \in \sem{\pi_\alpha(e)}(F_T)}
  $$
  $$
  \inferrule*[right=\anno{Union}]
    {t :: u \in \sem{e_1}(F_T) \cup \sem{e_2}(F_T)}
    {t :: u \in \sem{e_1 \cup e_2}(F_T)}
  \quad
  \inferrule*[right=\anno{Product}]
    {t_1 :: u_1 \in \sem{e_1}(F_T) \\ t_2 :: u_2 \in \sem{e_2}(F_T)}
    {(t_1 \otimes t_2) :: (u_1, u_2) \in \sem{e_1 \times e_2}(F_T)}
  $$
  $$
  \inferrule*[right=\anno{Intersect}]
    {t_1 :: u \in \sem{e_1}(F_T) \\ t_2 :: u \in \sem{e_2}(F_T)}
    {(t_1 \otimes t_2) :: u \in \sem{e_1 \cap e_2}(F_T)}
  $$
  $$
  \inferrule*[right=\anno{Natural-Join}]
    {t_1 :: (u, u_1) \in \sem{e_1}(F_T) \\ t_2 :: (u, u_2) \in \sem{e_2}(F_T)}
    {(t_1 \otimes t_2) :: (u, u_1, u_2) \in \sem{e_1 \bowtie e_2}(F_T)}
  $$
  $$
  \inferrule*[right=\anno{Diff-1}]
    {t :: u \in \sem{e_1}(F_T) \\ \sem{e_2}(F_T) \nvDash u}
    {t :: u \in \sem{e_1 - e_2}(F_T)}
  \quad
  \inferrule*[right=\anno{Diff-2}]
    {t_1 :: u \in \sem{e_1}(F_T) \\ t_2 :: u \in \sem{e_2}(F_T)}
    {(t_1 \otimes (\ominus~ t_2)) :: u \in \sem{e_1 - e_2}(F_T)}
  $$
  $$
  \inferrule*[right=\anno{AntiJoin-1}]
    {t :: (u_1, u_2) \in \sem{e_1}(F_T) \\ \sem{e_2}(F_T) \nvDash u_1}
    {t :: (u_1, u_2) \in \sem{e_1 \triangleright e_2}(F_T)}
  $$
  $$
  \inferrule*[right=\anno{AntiJoin-2}]
    {t_1 :: (u_1, u_2) \in \sem{e_1}(F_T) \\ t_2 :: u_1 \in \sem{e_2}(F_T)}
    {(t_1 \otimes (\ominus~ t_2)) :: (u_1, u_2) \in \sem{e_1 \triangleright e_2}(F_T)}
  $$
  $$
  \inferrule*[right=\anno{Aggregate}]
    {
      X_T \subseteq \sem{e}(F_T) \\
      \{ t_i :: u_i \}_{i=1}^n = X_T \\
      \{ \overline{t}_j :: \overline{u}_j \}_{j=1}^m = \sem{e}(F_T) - X_T \\
      u \in g(\{u_i\}_{i=1}^n)
    }
    {\textstyle (\bigotimes_{i=1}^n t_i) \otimes (\bigotimes_{j=1}^m (\ominus~ \overline{t}_j)) :: u \in \sem{\gamma_g(e)}(F_T)}
  $$
  $$
  \inferrule*[right=\anno{Group-By Aggregate}]
    {
      t' :: (u_g, u_g') \in \sem{e_1}(F_T) \\
      X_T \subseteq \sem{e_2}(F_T)
      \\\\
      \{ t_i :: (u_g, u_i) \}_{i=1}^n = X_T \\
      \{ \overline{t}_j :: (u_g, \overline{u}_j) \}_{j=1}^m = \sem{e_2}(F_T) - X_T \\
      u \in g(\{u_i\}_{i=1}^n)
    }
    {t' \otimes \textstyle (\bigotimes_{i=1}^n t_i) \otimes (\bigotimes_{j=1}^m (\ominus~ \overline{t}_j)) :: (u_g, u_g', u) \in \sem{\hat{\gamma}_g(e_1, e_2)}(F_T)}
  $$
  $$
  \inferrule*[right=\anno{Sample}]
    {
      \{ t_i :: u_i \}_{i=1}^n = \sem{e}(F_T) \\
      u_j \in \mu(\{\text{weight}(t_i) :: u_i\}_{i=1}^n)
    }
    { t_j :: u_j \in \sem{\psi_\mu(e)}(F_T) }
  $$
  $$
  \inferrule*[right=\anno{Group-By Sample}]
    {
      t' :: (u_g, u_g') \in \sem{e_1}(F_T) \\
      \{ t_i :: (u_g, u_i) \}_{i=1}^n \subseteq \sem{e}(F_T) \\
      u_j \in \mu(\{\text{weight}(t_i) :: u_i\}_{i=1}^n)
    }
    { (t' \otimes t_j) :: (u_g, u_g', u_j) \in \sem{\hat{\psi}_\mu(e_1, e_2)}(F_T) }
  $$
  \caption{Operational semantics of \ram.}
  \label{fig:full-sclram-semantics}
  \Description{}
\end{figure}

\begin{figure}
  \footnotesize
  \textbf{Rule semantics}
  \hfill
  \framebox[1.1\width]{$
    \left\langle . \right\rangle : \mathcal{U}_T \rightarrow \mathcal{U}_T, \quad
    \sem{r} : \mathcal{F}_T \rightarrow \mathcal{F}_T$}
  $$
  \inferrule*[right=\anno{Normalize}]
    {
      t_1 :: u, \dots, t_n :: u ~\text{are all tagged-tuples in}~ U_T \\
      t = \textstyle \bigoplus_{i=1}^n t_i \\
      \text{discard}(t) = \text{false}
    }
    { t :: u \in \left\langle U_T \right\rangle }
  $$
  $$
  \inferrule*[right=\anno{Rule-1}]
    {t^\old :: u \in \sem{p}(F_T) \\ \left\langle \sem{e}(F_T) \right\rangle \nvDash u}
    {t^\old :: p(u) \in \sem{p \leftarrow e}(F_T)}
  \quad
  \inferrule*[right=\anno{Rule-2}]
    {t^\new :: u \in \left\langle \sem{e}(F_T) \right\rangle \\ \sem{p}(F_T) \nvDash u}
    {t^\new :: p(u) \in \sem{p \leftarrow e}(F_T)}
  $$
  $$
  \inferrule*[right=\anno{Rule-3}]
    {t^\old :: u \in \sem{p}(F_T) \\ t^\new :: u \in \left\langle \sem{e}(F_T) \right\rangle}
    {(t^\old \oplus t^\new) :: p(u) \in \sem{p \leftarrow e}(F_T)}
  $$
  \textbf{Stratum and Program semantics}
  \hfill
  \framebox[1.1\width]{$\mathbf{lfp}^\circ : (\mathcal{F}_T \rightarrow \mathcal{F}_T) \rightarrow (\mathcal{F}_T \rightarrow \mathcal{F}_T), \quad \sem{s}, \sem{\overline{s}} : \mathcal{F}_T \rightarrow \mathcal{F}_T$}
  \begin{align*}
    \textsc{(Saturation)} &\quad
    F_T^\old \circeq F_T^\new ~\text{iff}~ \forall t^\new :: p(u) \in F_T^\new, \exists t^\old :: p(u) \in F_T^\old ~\text{such that}~ t^\old \oeq t^\new
    \\
    \textsc{(Fixpoint)} &\quad
    \mathbf{lfp}^\circ (h) = h \circ \dots \circ h = h^n ~\text{if there exists a minimum}~ n > 0,~\text{such that}~ h^{n}(F_T) \circeq h^{n+1}(F_T)
    \\
    \textsc{(Stratum)} &\quad
    \sem{s} = \mathbf{lfp}^\circ(\lambda F_T . (F_T - \textstyle \bigcup_{p \in P_s} F_T[p]) \cup (\textstyle \bigcup_{r \in s} \sem{r}(F_T)))
    \\
    \textsc{(Program)} &\quad
    \sem{\overline{s}} = \sem{s_n} \circ \dots \circ \sem{s_1}, ~\text{where}~ \overline{s} = s_1; \dots; s_n.
  \end{align*}
  \vspace{-0.15in}
  \caption{Operational semantics of \ram~ (Cont.)}
  \label{fig:full-sclram-semantics-2}
  \Description{}
\end{figure}

\subsection{Provenance Structures}

We provide detailed runtime analysis for each presented provenance structure in the following sections.
Additionally, we discuss the extended provenance \inlinecodesmall{diff-top-k-proofs-me} which can handle mutual exclusive facts.

\subsubsection{\inlinecodesmall{diff-max-min-prob}}
\paragraph{Runtime Analysis}
\begin{figure}[h]

\begin{minipage}{0.45\textwidth}
\centering
\footnotesize
\begin{tabular}{c|c|c}
    \textbf{Operation} & \textbf{Algorithm} & \textbf{Time Complexity} \\
    \hline
    $t_1 \oplus t_2$ & $\text{max}(t_1, t_2)$ & $O(1)$ \\
    $t_1 \otimes t_2$ & $\text{min}(t_1, t_2)$ & $O(1)$ \\
    $\ominus t$ & $\hat{1} - t$ & $O(1)$ \\
    $t_1 \oeq t_2$ & $t_1 == t_2$ & $O(1)$ \\
    $\rho(t)$ & $t$ & $O(1)$\\
\end{tabular}
\captionof{table}{\inlinecode{diff-max-min-prob} provenance structure run time analysis}
\label{tab:diff-max-min-prob-runtime}

\end{minipage}
\hfill
\begin{minipage}{0.45\textwidth}
\centering
\footnotesize
\begin{tabular}{c|c|c}
    \textbf{Operation} & \textbf{Algorithm} & \textbf{Time Complexity} \\
 \hline
    $t_1 \oplus t_2$ & $clamp(t_1 + t_2)$ & $O(n)$ \\
    $t_1 \otimes t_2$ & $t_1 \cdot t_2$ & $O(n)$ \\
    $\ominus t$ & $\hat{1} - t$ & $O(n)$ \\
    $t_1 \oeq t_2$ & true & $O(1)$ \\
    $\rho(t)$ & $t$ & $O(1)$\\
\end{tabular}
\captionof{table}{\inlinecode{diff-add-mult-prob} provenance structure run time analysis}
\label{tab:diff-add-mult-prob-runtime}

\end{minipage}

\begin{minipage}{1.0\textwidth}
\centering
\footnotesize
\begin{tabular}{c|c|c}
    \textbf{Operation} & \textbf{Algorithm} & \textbf{Time Complexity} \\
    \hline
    $t_1 \oplus t_2$ & $\text{top}_k(t_1 \cup t_2)$ & $O(nk)$ \\
    $t_1 \otimes t_2$ & $\text{top}_k(\{~ \eta \sep (\eta_1, \eta_2) \in t_1 \times t_2, \eta = \eta_1 \cup \eta_2, \eta ~\text{no conflict} ~\})$ & $O(n^2k^2)$ \\
    $\ominus t$ & $\text{top}_k(\text{cnf2dnf}(\{ \{\neg \nu \sep \nu \in \eta \} \sep \eta \in t \}$ & $O(2^n)$ \\
    $t_1 \oeq t_2$ & $t_1 == t_2$ & $O(nk)$ \\
    $\rho(t)$ & WMC$(t, \Gamma)$ & $O(2^n)$
\end{tabular}
\captionof{table}{\inlinecode{diff-top-k-proofs} provenance structure run time analysis. We assume the term number for each tag is of $O(n)$, and the number of clauses is of $O(k)$. Note that we show the time complexity for the naive implementation of cnf2dnf and WMC algorithms.}
\label{tab:diff-top-k-proofs-runtime}

\end{minipage}

\end{figure}
We showcase the time complexity for each operator in the max-min-prob provenance structure in \tabref{tab:diff-max-min-prob-runtime}.
We next present our optimized counting algorithm in \figref{alg:max-min-prob-aggr} which performs counting over a set of \inlinecodesmall{max-min-prob} tagged-tuples.
Note that we only showed the algorithm for \inlinecodesmall{mmp} for simplicity.
But it easily extends to \inlinecodesmall{dmmp}.
It can be easily shown that its runtime complexity is $O(n \log(n))$.

\begin{algorithm}
\caption{Counting over max-min-prob tagged tuples}
\label{alg:max-min-prob-aggr}
\KwData{$U_\texttt{mmp} = \{ t_1 :: u_1, t_2 :: u_2, \dots, t_n :: u_n \}$: $\mathcal{U}_\texttt{mmp}$, set of tagged-tuples}
\KwResult{$U_\texttt{mmp}'$: $\mathcal{U}_\texttt{mmp}$}
\tcc{sort all positive tuples according to their tags from small to large. O(nlog(n))}
$\textbf{t}^\text{pos} = sorted([t_i \sep i = 1 \dots n])$\;
$\textbf{t}^\text{neg} = [1 - t^\text{pos}_{n - i + 1} \sep i = 1 \dots n]$\;

\tcc{Iterate through all possible partitions between positive and negative tags. O(n)}
$U_\texttt{mmp}'$ = $\{t^\text{neg}_n :: 0, t^\text{pos}_1 :: n\}$ \;
\For{$i = 1 \dots (n - 1)$}{
  Add $\text{min}(t^\text{pos}_{i + 1}, t^\text{neg}_{i}) :: (n - i)$ to $U_\texttt{mmp}'$\;
}
\Return $U_\texttt{mmp}'$
\end{algorithm}

\subsubsection{\inlinecodesmall{diff-add-mult-prob}}
\paragraph{Runtime Analysis}
We showcase the time complexity for the diff-add-mult-prob provenance structure in \tabref{tab:diff-add-mult-prob-runtime}. We denote the number of total variables in the input as $n$.
Since we are performing calculations over the dual-numbers, the time complexity is decided by the gradient calculation over vectors of dimension $n$.

\subsubsection{\inlinecodesmall{diff-top-k-proofs}}

\paragraph{Runtime Analysis}
We showcase the time complexity for the diff-top-k-proofs provenance structure in \tabref{tab:diff-top-k-proofs-runtime}. Note that we show the runtime analysis for the naive implementation of the cnf2dnf function and weighted model counting process.
We leave the optimizations for future work.

\subsubsection{\inlinecodesmall{diff-top-k-proofs-me}}

\paragraph{Support for Mutually Exclusive Facts}

We keep the original \inlinecodesmall{diff-top-k-proofs} and extend it to support mutal exclusive tags (hence the name \inlinecodesmall{-me}).
Most of the operations remains, and we start by giving it a new input tag space:
$$I_{\inlinecodesmall{dtkp-me}} = [0, 1] \times \texttt{Option<}\mathbb{N}\texttt{>}.$$
The probability remains the same, and the second part is an optional identifier of one mutual exclusion.
Not providing that means the fact does not belong to a mutual exclusion.
If presented, we denote such a tag
$$t_i = (r_i, \texttt{Some}(\textbf{me}_i))$$

\begin{example}
  Consider the sum2 task, we have two distributions specified by
  \begin{lstlisting}[language=scallop,numbers=none,xleftmargin=.05\textwidth]
rel digit = {0.90::(A, 0); 0.01::(A, 1); ...; 0.02::(A, 9)}
rel digit = {0.01::(B, 0); 0.91::(B, 1); ...; 0.01::(B, 9)}
  \end{lstlisting}
  Then we would have the input facts being
  \[
  \sem{\texttt{digit}}(F_{\texttt{Option<}I\texttt{>}})
  =
  \left\{
  \begin{array}{rcl}
    (0.90, \inlinecodesmall{Some}(1)) & :: & (A, 0), \\
    (0.01, \inlinecodesmall{Some}(1)) & :: & (A, 1), \\
    & \vdots & \\
    (0.02, \inlinecodesmall{Some}(1)) & :: & (A, 9), \\ \hline
    (0.01, \inlinecodesmall{Some}(2)) & :: & (B, 0), \\
    (0.91, \inlinecodesmall{Some}(2)) & :: & (B, 1), \\
    & \vdots & \\
    (0.01, \inlinecodesmall{Some}(2)) & :: & (B, 9) \\
  \end{array}
  \right\}.
  \]
  There are two $\textbf{me}$-s, namely $1$ and $2$, suggesting there are two mutual exclusions.
  The first 10 facts for digit $A$ belong to mutual exclusion $1$, and the next 10 for digit $B$ belong to mutual exclusion $2$.
\end{example}

Putting it in action, two facts $f_i$ and $f_j$ with the same \textbf{me} ($\textbf{me}_i = \textbf{me}_j$) cannot appear together in the same proof.
This is enforced inside of the conflict check during \inlinecodesmall{dtkp}'s $\wedge_k$ and $\neg_k$ operations.
If the proof violates the mutual exclusive assumption, it will be deemed invalid proof and discarded.
For example, executing the following rule
\begin{lstlisting}[language=scallop,numbers=none,xleftmargin=.05\textwidth]
rel not_possible() = digit(A, x), digit(A, y), x != y
\end{lstlisting}
shall yield no valid output due to this mechanism, i.e., the internal tag associated with \inlinecodesmall{not\_possible()} is going to be $\emptyset$ and be early-removed from the computation.

\section{Evaluation}
\label{app:evaluation}

\subsection{MNIST-R}
\paragraph{Experimental Setup}
The total MNIST\cite{lecun1998mnist} dataset contains 60K handwritten digits.
According to the number of digits used in each task, we further split the MNIST dataset into downstream task data points.
For example, we have 30K training pairs for sum-2, 20K training pairs for sum-3, and 15K training pairs for sum-4.
As the counting task consumes $8$ images per datapoint, we have 7.5K training pairs for the counting tasks.
The test and training splits follow the original MNIST task setup.
Our neural module consists of a 2-layer CNN connected by a 2-layer MLP, with a hidden layer size of $1024$.
We trained the \ours~ module $10$ epochs on the training dataset, the learning rate is set to $0.001$, and the batch size is $64$, with a BCE-loss closing the training loop.

\paragraph{\ours~ code}
We present the \ours~ code that is used in the MNIST-R benchmark in \figref{fig:mnist-r-scl-code}.

\begin{figure}
    \input{figures/scl_codes/mnist/sum-2}
    \hfill
    \input{figures/scl_codes/mnist/sum-3}
    \hfill
    \input{figures/scl_codes/mnist/sum-4}
    \hfill
    \input{figures/scl_codes/mnist/less-than}
    \hfill
    \input{figures/scl_codes/mnist/not-3-or-4}
    \hfill
    \input{figures/scl_codes/mnist/count-3}
    \hfill
    \input{figures/scl_codes/mnist/count-3-or-4}
    \caption{\ours~ code for MNIST-R.}
    \label{fig:mnist-r-scl-code}
\end{figure}

\subsection{Hand Written Formula}
\begin{figure}
    \centering
    \begin{lstlisting}[language=scallop,numbers=left,xleftmargin=.05\textwidth]
type symbol(index: usize, symbol: String)
type length(n: usize)

rel digit = {"0", "1", "2", "3", "4", "5", "6", "7", "8", "9"}

type term(value: f32, begin: usize, end: usize)
rel term(x as f32, b, b + 1) = symbol(b, x) and digit(x)

type mult_div(value: f32, begin: usize, end: usize)
rel mult_div(x, b, r) = term(x, b, r)
rel mult_div(x * y, b, e) = mult_div(x, b, m) and symbol(m, "*") and
                            term(y, m + 1, e)
rel mult_div(x / y, b, e) = mult_div(x, b, m) and symbol(m, "/") and
                            term(y, m + 1, e)

type add_minus(value: f32, begin: usize, end: usize)
rel add_minus(x, b, r) = mult_div(x, b, r)
rel add_minus(x + y, b, e) = add_minus(x, b, m) and symbol(m, "+") and
                             mult_div(y, m + 1, e)
rel add_minus(x - y, b, e) = add_minus(x, b, m) and symbol(m, "-") and
                             mult_div(y, m + 1, e)

type result(value: f32)
rel result(y) = add_minus(y, 0, l) and length(l)

query result
    \end{lstlisting}
    \caption{\ours~ code for HWF.}
    \label{fig:hwf-code}
    \Description{}
\end{figure}

\paragraph{Experimental Setup}
The HWF~\cite{li2020closed} dataset takes in a sequence of images, where each image can be either a digit or an arithmetic symbol, and aims to calculate the outcome of the formula.
We apply the same CNN-based symbol recognition network as for the MNIST-based tasks, with the only difference being that the model now performs a 14-way classification.
We then generate a probabilistic database containing two relations:
1) \inlinecodesmall{length(usize)}, a non-probabilistic relation storing the length of the formula, and
2) \inlinecodesmall{symbol(id: usize, $\sigma$: String)}, a probabilistic relation mapping index IDs to symbols $\sigma \in \Sigma$.
For the reasoning component, we write a formula parser in \ours~ (as shown in \ref{fig:hwf-code}).
Additionally, \ours~ is used to perform calculations based on the parsed abstract syntax tree (AST) to obtain the final rational number output.
The training loop is closed by a BCE loss at the end, forming an end-to-end training pipeline where gradients can back-propagate all the way from the output to the symbol recognition network.
To prune the enormous output domain if we enumerate every possible parse tree, we sample only a subset of symbol predictions to be sent to \ours.
There are 10K images answer pairs for the training dataset.
We set the batch size to 16, the learning rate to 0.0001, and the training epochs to 100.
To enhance the learning efficiency, we adopt a sampling strategy, which only preserves the $7$ most likely classification result for each image classification task, and sends it to \ours.
\subsection{Pathfinder}

\paragraph{Problem Definition}
\begin{figure}
    \centering
    \footnotesize
    \begin{tabular}[c]{c||m{0.6in}|m{0.6in}|m{0.6in}}
        \toprule \hline
        & Easy & Medium & Hard \\ \hline
        \begin{tabular}{c}Positive \\ (Connected)\end{tabular} & \includegraphics[height=0.6in,valign=b]{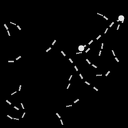} & \includegraphics[height=0.6in,valign=b]{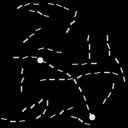} & \includegraphics[height=0.6in,valign=b]{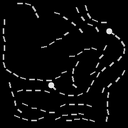} \\ \hline \hline
        \begin{tabular}{c}Negative \\ (Disconnected)\end{tabular} & \includegraphics[height=0.6in,valign=b]{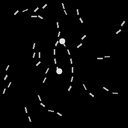} & \includegraphics[height=0.6in,valign=b]{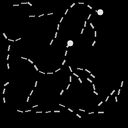} & \includegraphics[height=0.6in,valign=b]{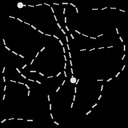} \\ \hline
        \bottomrule
    \end{tabular}
    \caption{Examples of positive and negative for each difficulty level.}
    \label{tab:pathfinder-example}
    \Description{}
\end{figure}

\begin{figure}
    \centering
    \begin{lstlisting}[language=scallop,numbers=left,xleftmargin=.05\textwidth]
// Input from neural networks
type dash(i8, i8)
type dot(i8)

// Connectivity check
rel path(x, y) = dash(x, y) or path(x, z) and dash(z, y)
rel connected() = dot(x), dot(y), path(x, y), x != y
    \end{lstlisting}
    \caption{\ours~ code for Pathfinder.}
    \label{fig:pathfinder-code}
\end{figure}

Pathfinder is a task that initially set out to test the long-range reasoning ability of neural networks.
It is adopted into a benchmark called Long Range Arena (LRA) \cite{tay2020longrange} to evaluate the long-range reasoning ability of neural networks, or more specifically the Transformers~\citep{transformer}.
In this task, we are provided with images with dark backgrounds and two white dots possibly connected by dashes.
The model is then required to make a binary prediction on whether the two dots are connected or not.
There are multiple versions of the dataset: one consisting of only $32 \times 32$ images (Path) and the other one with $128 \times 128$ (Path-X).
Both versions have 3 difficulties, easy, normal, and hard, depending on the length of the dashes connecting the two dots -- longer means harder.
Table \ref{tab:pathfinder-example} shows the 3 difficulties each with one positive and one negative sample from Path-X.

\paragraph{Architecture}

Going to the extreme of symbolism one would use the individual symbol to represent each small dash and dot.
However, doing so would mean that the perception network, presumably a CNN, needs to segment the image precisely to extract all dashes and dots as entities.
At the same time, the reasoning module bears an extra burden when processing individual connectivities.
Instead, we employ a simple grid-based connectivity graph:
each node represents a conceptual ``dot'' and each edge represents a conceptual ``dash''.
We ask neural networks to take in the image and output a feature vector with the size of the number of dots plus the number of edges.
Each logit in this vector is then treated as a probability of that ``dot'' or ``dash''.
The \ours~ program we use for this task, as shown in  \figref{fig:pathfinder-code}, is simply a transitive closure that checks whether there are two distinct dots connected by dashes.

\paragraph{Experimental Setup}
There are 600K training data points for both PATH and PATH-X, which are images of $32 \times 32$ and $128 \times 128$ pixels respectively, and the y-label represents whether there exists a path that connects the two dots on the image.
We recognize the ``dash'' and ``dots'' relations through a 4-layer CNN, and the local connectivity is recognized through a 2-layer MLP classifier.
We set the learning rate to $0.0001$, the batch size to $64$, and the number of epochs to $100$.

\subsection{PacMan-Maze}
\begin{figure}
    \centering
    \begin{lstlisting}[language=scallop,numbers=left,xleftmargin=.05\textwidth]
// Input from neural networks
type grid_node(x: usize, y: usize) // 0.99 for each grid cell
type actor(x: usize, y: usize)
type goal(x: usize, y: usize)
type enemy(x: usize, y: usize)

// Basic connectivity
const UP = 0, RIGHT = 1, DOWN = 2, LEFT = 3
rel safe_node(x, y) = grid_node(x, y), not enemy(x, y)
rel edge(x, y, x, yp, UP) = safe_node(x, y), safe_node(x, yp), yp == y + 1
rel edge(x, y, xp, y, RIGHT) = safe_node(x, y), safe_node(xp, y), xp == x + 1
rel edge(x, y, x, yp, DOWN) = safe_node(x, y), safe_node(x, yp), yp == y - 1
rel edge(x, y, xp, y, LEFT) = safe_node(x, y), safe_node(xp, y), xp == x - 1

// Get the next position
rel next_pos(xp, yp, a) = actor(x, y), edge(x, y, xp, yp, a)

// Path for connectivity; will condition on no enemy on the path
rel path(x, y, x, y) = next_pos(x, y, _)
rel path(x1, y1, x3, y3) = path(x1, y1, x2, y2), edge(x2, y2, x3, y3, _)

// Get the next action
rel next_action(a) = next_pos(x, y, a), goal(gx, gy), path(x, y, gx, gy)

// Constraint violation
type too_many_goal(), too_many_actor(), too_many_enemy()
rel too_many_goal() = n := count(x, y: goal(x, y)), n > 1
rel too_many_actor() = n := count(x, y: actor(x, y)), n > 1
rel too_many_enemy() = n := count(x, y: enemy(x, y)), n > 5
rel violation() = too_many_goal() or too_many_enemy() or too_many_actor()
    \end{lstlisting}
    \caption{\ours~ code for PacMan-Maze.}
    \label{fig:pacman-maze-code}
    \Description{}
\end{figure}

PacMan-Maze is a reinforcement learning based planning application, which aims to lead the actor to the flag without running into any enemy. We show the \ours~ code for this task in \figref{fig:pacman-maze-code}.
In addition to the code shown in \secref{sec:overview}, we added additional ``integrity constraint violation'' terms that need to be minimized during learning.

\paragraph{Experimental Setup}
We set the size of the maze to be a $5 \times 5$ matrix, and the maximum number of enemies that exists in the arena to be $5$. In the training process, we set the batch size to be $24$, the memory replay buffer size to be $3000$, and the maximum number of actions in an episode to be $30$. We update the target net every 10 epochs, we set the exploration rate to be $0.9$, its falloff to be $0.98$, and the learning rate is set to $0.0001$.

\subsection{CLUTRR}
\begin{figure}
    \centering
    \begin{lstlisting}[language=scallop,numbers=left,xleftmargin=.05\textwidth]
// Define kinships belong to the Relation type
type Relation = usize

//  Input from neural networks
type question(sub: String, obj: String)
type kinship(rela: Relation, sub: String, obj: String)
type composition(Relation, Relation, Relation)

// Rules to derive the final answer
rel kinship(r3, x, z) = composition(r1, r2, r3), kinship(r1, x, y),
                        kinship(r2, y, z), x != z
rel answer(r) = question(s, o), kinship(r, s, o)
    \end{lstlisting}
    \caption{\ours~ code for CLUTRR.}
    \label{fig:clutrr-code}
\end{figure}

With CLUTRR \cite{sinha2019clutrr} we step into the domain of NLP on kinship reasoning.
Each data point in the CLUTRR dataset consists of a natural language passage, a query, and its answer.
The passage tells a story about a family's daily activities;
the query is a tuple of two names, denoting the target pair of persons that we want to predict the relation for;
the answer is among the 20 basic relations such as ``mother'', ``uncle'', and ``daughter-in-law''.
The dataset is further categorized into 10 difficulty levels according to the number of facts ($k$) required to derive the final result.
Note that the knowledge base for kinship compositionality, such as "father's father is grandfather" is not included in the dataset.
We, therefore, designed three experiments to help alleviate the issue of missing knowledge base, a) learning with manually specified rules, b) rule learning, and c) learning without rules.
We showcase our \ours~ code in \figref{fig:clutrr-code}.

\paragraph{Learn with Manually Specified Rules}

A straightforward strategy to overcome the missing required database issue, is to manually add it.
We created an external knowledge base with 92 kinship composition triplets and 3 rules.
To define how two known relations can be connected to derive the third relation, we design a high-order predicate, \inlinecodesmall{composition}.
As an example, the rule ``father's mother is grandmother'' is encoded as \inlinecodesmall{composition(FATHER, MOTHER, GRANDMOTHER)}.
The full \ours~ program connecting the knowledge base into the reasoning process is shown in \figref{fig:clutrr-code}
Given the knowledge base, context, and question-answer pairs, we are ready to perform learning over the new dataset.
As the context has varies lengths, we first separate the context into multiple windows to ensure the perception model processes similar-length-context across the dataset.
Then a large language model, such as RoBERTa \cite{liu2019roberta}, is adopted to extract relations from these windowed texts.
The kinship extracted from the context goes into \inlinecodesmall{kinship(rela, sub, obj)}.
For example, \inlinecodesmall{kinship(SON, "Alice", "Bob")}
indicates that Bob is Alice's son.
The query, on the other hand, is stored in the relation \inlinecodesmall{question(sub, obj)}.
Next, we will combine the predicted relations from all the context windows into one probabilistic database, together with the knowledge base, and the query to perform logical reasoning.
Last, we compare the probabilistic query result with the ground truth with BCE loss.

\paragraph{Rule Learning}
Specifying 92 composition rules manually could be time consuming.
With \ours, it is possible to do this in a smarter way.
Since there are 20 basic kinship relations, we have a finite space containing $20 ^ 3 = 8\text{K}$ possible composition facts.
We treat all 8K composition facts as probabilistic and to be learnt, and therefore we store them inside a tensor of size $20 \times 20 \times 20$.
We can initialize this tensor randomly but with a low maximum weight (i.e. $0.1$).
During training, we will allow gradients to be back-propagated to this composition tensor so that these weights can also be learnt.
In this case, the user does not specify any composition fact manually and everything is learnt on the fly.
Note that, 8K composition facts can slow down the training time.
Therefore, we use multinomial sampling to pick 150 composition facts for training.
During testing, we simply select the top 150 for inference.

\paragraph{Learn without Rules} We can also jointly extract entity relations and learn the composition rules without the human specified rules. This is a more challenging task where one needs to figure out the way to combine inferred relations and derive the desired answer without intermediate supervision.

\paragraph{Experimental Setup}
We have 10K training data points, which are triplets of passage, query, and answers.
We use RoBERTa as the backbone language model.
We further train a 2-layer MLP-based relation extractor which takes in the embedding of three components, a global embedding of the passage, and the embedding of the queried subject and the object, and returns the result for a 21-way classification.
We set the batch size to $32$, the learning rate to
$0.00001$, and train the models for $100$ epochs.

\subsection{Mugen}
\begin{figure}
    \centering
    \begin{lstlisting}[language=scallop,numbers=left,xleftmargin=.05\textwidth]
// Input from neural networks
type action(usize, String)
type expr(usize, String)
type expr_start(usize)
type expr_end(usize)
type action_start(usize)
type action_end(usize)

type match_single(usize, usize, usize)
type match_sub(usize, usize, usize, usize)

// Check whether does a subsection of text specification matches the
// video content
rel match_single(tid, vid, vid + 1) = expr(tid, a), action(vid, a)
rel match_sub(tid, tid, vid_start, vid_end) =
    match_single(tid, vid_start, vid_end)
rel match_sub(tid, tid, vid_start, vid_end) =
    match_sub(tid, tid, vid_start, vid_mid),
    match_single(tid, vid_mid, vid_end)
rel match_sub(tid_start, tid_end, vid_start, vid_end) =
    match_sub(tid_start, tid_end - 1, vid_start, vid_mid),
    match_single(tid_end, vid_mid, vid_end)

// Check whether does the whole text specification matches the video content
rel match() = expr_start(tid_start),
    expr_end(tid_end), action_start(vid_start), action_end(vid_end),
    match_sub(tid_start, tid_end, vid_start, vid_end)

// Integrity violation when too many identical text expressions
// occurs consecutively
rel too_many_consecutive_expr() = expr(tid, a),
    expr(tid + 1, a), expr(tid + 2, a), expr(tid + 3, a)
    \end{lstlisting}
    \caption{\ours~ code for Mugen.}
    \label{fig:mugen-code}
    \Description{}
\end{figure}

\paragraph{Experimental Setup}
We sample 1K Mugen\cite{hayes2022mugen} video and text pairs as the training dataset, and another 1K for testing.
Our neural module consists of the S3D image embedding and the distillBert text embedding.
Then we pass the embeddings through a 2-layer MLP, with a hidden layer size of 256.
We trained the \ours~ module 1000 epochs on the training dataset, the learning rate is set to 0.0001, and the batch size is 3, with BCE-loss closing the training loop.
The \ours~ code is shown in \figref{fig:mugen-code}

\subsection{CLEVR}
\begin{figure}
    \centering
    \begin{lstlisting}[language=scallop,numbers=left,xleftmargin=.05\textwidth]
// yes/no question
rel eval_yn(e, x > y) = greater_than_expr(e, a, b),
                    eval_num(a, x), eval_num(b, y)
rel eval_yn(e, x < y) = less_than_expr(e, a, b),
                    eval_num(a, x), eval_num(b, y)
rel eval_yn(e, x == y) = equal_expr(e, a, b),
                    eval_num(a, x), eval_num(b, y)
rel eval_yn(e, x == y) = equal_color_expr(e, a, b),
                    eval_query(a, x), eval_query(b, y)
rel eval_yn(e, x == y) = equal_material_expr(e, a, b),
                    eval_query(a, x), eval_query(b, y)
rel eval_yn(e, x == y) = equal_shape_expr(e, a, b),
                    eval_query(a, x), eval_query(b, y)
rel eval_yn(e, x == y) = equal_size_expr(e, a, b),
                    eval_query(a, x), eval_query(b, y)
rel eval_yn(e, b) = b := exists(o: eval_objs(f, o)
                    where e: exists_expr(e, f))

// count number of objects
rel eval_num(e, n) = n := count(o: eval_objs(f, o) where e: count_expr(e, f))

// objects filter
rel eval_objs(e, o) = scene_expr(e), obj(o)
rel eval_objs(e, o) = unique_expr(e, f), eval_objs(f, o)
rel eval_objs(e, o) = filter_size_expr(e, f, s), eval_objs(f, o), size(o, s)
rel eval_objs(e, o) = filter_color_expr(e, f, c), eval_objs(f, o), color(o, c)
rel eval_objs(e, o) = filter_material_expr(e, f, m), eval_objs(f, o), material(o, m)
rel eval_objs(e, o) = filter_shape_expr(e, f, s), eval_objs(f, o), shape(o, s)
rel eval_objs(e, o) = and_expr(e, f1, f2), eval_objs(f1, o), eval_objs(f2, o)
rel eval_objs(e, o) = or_expr(e, f1, f2) and (eval_objs(f1, o) or eval_objs(f2, o))

// same attribute
rel eval_objs(e, o) = same_size_expr(e, f),
            eval_objs(f, p), size(p, c), size(o, c), o != p
rel eval_objs(e, o) = same_color_expr(e, f),
            eval_objs(f, p), color(p, c), color(o, c), o != p
rel eval_objs(e, o) = same_material_expr(e, f),
            eval_objs(f, p), material(p, m), material(o, m), o != p
rel eval_objs(e, o) = same_shape_expr(e, f),
            eval_objs(f, p), shape(p, s), shape(o, s), o != p

// relation
rel relate("right", p, o) = relate("left", o, p)
rel relate("front", p, o) = relate("behind", o, p)
rel eval_objs(e, o) = relate_expr(e, f, r),
                    eval_objs(f, p), relate(r, p, o), o != p

// eval query
rel eval_query(e, s) = query_size_expr(e, f), eval_objs(f, o), size(o, s)
rel eval_query(e, c) = query_color_expr(e, f), eval_objs(f, o), color(o, c)
rel eval_query(e, m) = query_material_expr(e, f), eval_objs(f, o), material(o, m)
rel eval_query(e, s) = query_shape_expr(e, f), eval_objs(f, o), shape(o, s)
\end{lstlisting}
\end{figure}

\begin{figure}
\begin{lstlisting}[language=scallop,numbers=left,xleftmargin=.05\textwidth, firstnumber=54]
// Final result
rel result(y as String) = root_expr(e), eval_yn(e, y)
rel result(y as String) = root_expr(e), eval_num(e, y)
rel result(y) = root_expr(e), eval_query(e, y)
    \end{lstlisting}
    \caption{\ours~ code for CLEVR.}
    \label{fig:clevr-code}
\end{figure}

CLEVR \cite{clevr} stands for \textbf{C}ompositional \textbf{L}anguage and \textbf{E}lementary \textbf{V}isual \textbf{R}easoning,
is a synthetic dataset testing model's ability to perform \textit{Visual Question Answering} (VQA).
Each data point of the dataset contains an image and a question-answer pair with regard to the image.
The images are randomly generated from a dictionary of 3 shapes, 8 colors, 2 materials, and 2 sizes and the programmatic queries are also randomly generated.
Given a rendered image with simple objects such as cubes and spheres, we aim to answer a question about the image.

\paragraph{Architecture}
The object feature vectors are obtained by a 4-layer convolutional neural network.
The feature vectors are then passed to four 2-layer MLP classifiers and another 3-layer MLP classifier to obtain the attributes and the spacial relations respectively.
The learning rate is set to $0.0001$, the batch size is $32$, and the latent dimension is $1024$. We show our \ours~ code in \figref{fig:clevr-code}.

\end{document}